\documentclass[draft]{agujournal2018}

\draftfalse

\journalname{JGR: Space Physics}

\begin{document}

%
%


\title{Validation of the neutron monitor yield function using data from AMS-02 experiment, 2011\,--\,2017}

%
%




\authors{Sergey A. Koldobskiy \affil{1, 2},
    Veronica Bindi \affil{3},
    Claudio Corti \affil{3},
	Gennady A. Kovaltsov \affil{4},
	Ilya G. Usoskin \affil{1}
}

\affiliation{1}{University of Oulu, Finland}
\affiliation{2}{National Research Nuclear University ``MEPhI'', Moscow, Russia}
\affiliation{3}{University of Hawaii at Manoa, Honolulu, USA}
\affiliation{4}{Ioffe Physical-Technical Institute, St. Petersburg, Russia}




\correspondingauthor{Ilya Usoskin}{ilya.usoskin@oulu.fi}




\begin{keypoints}
\item A direct validation of the neutron monitor (NM) yield function is performed using the proton and helium spectra measured by AMS-02
 for the period of 2011\,--\,2017.
\item The NM yield function by \citet{mishev13} is fully validated, while others need revisions.
\item The use of the force-field approximation to fit the measured proton spectra is validated within $\pm 10$\,\%.
\end{keypoints}

%
%


\begin{abstract}
The newly published spectra of protons and helium over time directly measured in space by the AMS-02 experiment for the period 2011\,--\,2017
 provide a unique opportunity to calibrate ground-based neutron monitors (NMs).
Here, calibration of several stable sea-level NMs (Inuvik, Apatity, Oulu, Newark, Moscow, Hermanus, Athens) was performed
 using these spectra.
Four modern NM yield functions were verified: Mi13 \citep{mishev13}, Ma16 \citep{mangeard2}, CM12 \citep{caballero12} and CD00 \citep{clem00},
 on the basis of the cosmic-ray spectra measured by AMS-02.
The Mi13 yield function was found to realistically represent the NM response to galactic cosmic rays.
CM12 yield function leads to a small skew in the solar cycle dependence of the scaling factor.
In contrast, Ma16 and CD00 yield functions tend to overestimate the NM sensitivity to low-rigidity ($<$10 GV) cosmic rays.
This effect may be important for an analysis of ground level enhancements, leading to a potential
 underestimate of fluxes of solar energetic particles as based on NM data.
The Mi13 yield function is recommended for quantitative analyses of NM data, especially for ground-level enhancements.
The validity the force-field approximation was studied, and it was found that it fits well the directly measured proton spectra,
 within a few \% for periods of low to moderate activity and up to $\approx 10$\,\% for active periods.
The results of this work strengthen and validate the method of the cosmic-ray variability analysis based on the NM data and
 yield-function formalism, and improves its accuracy.
\end{abstract}

%
%
%


%
%

%


%
%
%
%

\section{Introduction}
\label{intro}

Cosmic rays form a permanent but variable radiation environment at Earth, being in particular the main agent ionizing atmosphere
 at low and moderate heights \citep[e.g.,][]{vainio09}.
The most important ones are galactic cosmic rays (GCRs) which are always present in the vicinity of the Earth
 and can be very energetic, up to $10^{20}$ eV, although the bulk has energy of several GeV.
Sometimes, eruptive solar events (flares or coronal mass ejections) can accelerate energetic particles
  leading to sporadic solar energetic particle (SEP) events with greatly enhanced fluxes for hours or days.
Here we focus primarily on the GCR measurements.
The GCR flux varies in time as a result of solar modulation in the heliosphere, that consists of diffusion, convection,
 and adiabatic cooling in the radially expanding solar wind, as well as drifts \citep[see, e.g., a review by][]{potgieterLR}.

The main instrument to measure long-term cosmic-ray variability is the world-wide network of standard ground-based neutron monitors (NMs),
 which is in continuous operation since the 1950s and provides the reference dataset for GCR modulation for nearly 70 years
 \citep{belov00,simpson00,shea_SSR_00}.
Since NM is an energy-integrating detector, which uses the whole atmosphere above it as a moderator, it is not trivial to
 relate the count rate of a NM to the flux or energy spectrum of GCRs at the top of the atmosphere.
One needs to know the so-called NM yield function, \textit{i.e.} the response of a standard NM to a unit flux of cosmic rays
 with the given energy.
Yield functions are calculated either theoretically, using a numerical simulation of the nucleonic cascade caused by
 energetic cosmic rays in the Earth's atmosphere \citep[e.g.,][]{clem00}, or semi-empirically, based on an NM
 latitudinal survey  \citep{caballero12}.
Yield functions differ quite a bit between each other (see Section~\ref{Sec:YF}), and until recently there was no direct way to
 validate and calibrate them \citep[see][for details]{dorman04}.

The situation has been changed recently when precise space-borne spectrometers had been launched to measure
 cosmic-ray spectra in situ, outside the Earth's atmosphere: the Payload for Antimatter Matter Exploration and Light-nuclei Astrophysics
 (PAMELA) instrument was in operation during 2006\,--\,2016 \citep{adriani14,adriani17}, while the Alpha Magnetic Spectrometer (AMS-02)
 is in continuous operation since 2011 \citep{aguilar_AMS_18}.
An attempt to directly calibrate the NM yield functions was performed recently \citep{usoskin_gil_17,koldobsky18} using proton spectra
 measured by PAMELA with nearly monthly resolution for the period 2006\,--\,2014.
Although a discrepancy between the modelled and measured count rates of several NMs was found, the reason was unclear.
It is likely because of the ambiguity related to the contribution
 of heavier-than-proton species of cosmic rays, since only proton spectra were available from PAMELA measurements.
Recently, measured spectra of protons and helium nuclei were published by the AMS-02 collaboration \citep{aguilar_AMS_18}
 for individual 27-day periods (Bartels rotations, BRs) between May 2011 and May 2017.
This new data makes it possible to perform a full validation and calibration of the NM yield functions.
This forms the main topic of this work.
In addition, we assess the validity of the widely used force-field approximation \citep{caballero04,usoskin_Phi_05}
 to describe the modulated GCR proton spectrum based on the AMS-02 data.

The paper is organized as follows:
selection and sources of the data used are described in Section~\ref{data}, calibration and validation of the NM yield functions
 are presented in Section~\ref{scale_factor}, validation of the force-field parameterization is discussed in Section~\ref{phi},
 and conclusions are summarized in Section~\ref{conc}.

\section{Data selection}
\label{data}
Recently published data of the AMS-02 collaboration on proton and helium energy/rigidity spectra were used in this work.
AMS is a high-precision magnetic spectrometer installed on the International Space Station (ISS) in May 2011, capable of measuring GCR
 fluxes in the rigidity range 1 GV -- 3 TV with uncertainties well below 10\%  \citep{aguilar_AMS_13}.
The low orbit of ISS is inside the Earth's magnetosphere, but thanks to its significant inclination of $\approx 52^\circ$ to
 the equator, it allows one to measure particles with rigidity as low as $\approx 1$ GV, which is sufficient for the present
 study, since the relative contribution of $<1$ GV particles to the count rate of a polar NM is less than $10^{-4}$
 \citep{asvestari_JGR_17}.
We used the proton and helium spectra measured from May 2011 through May 2017 for 79 BRs
 2426\,--\,2506\footnote{BRs 2472 and 2473 are missing in the data.} \citep{aguilar_AMS_18}
 as available at the NASA Space Physics Data Facility (SPDF) Database (see Acknowledgements).
The spectra measured for individual BRs are limited to rigidities below 60.3 GV.
For the higher rigidity/energy range, not affected by the heliospheric modulation, we used measured spectra of protons and helium
 integrated over the 30-month period from May 2011 through November 2013 \citep{aguilar15,aguilar_he_15}.
We also used proton spectra measured during the PAMELA experiment, AMS-01, and several balloon-borne flights.
Tabulated data were adopted from the Space Science Data Center (SSDC) database.
We also used the count rates (normalized to one counter) of several nearly sea-level stably running NMs (see Table~\ref{Tab:NM}) as
 collected from the NMDB (Neutron Monitor Database) project as well as dedicated web-resources for
  Oulu (\url{http://cosmicrays.oulu.fi/}) and Apatity (\url{http://pgia.ru/CosmicRay/}) NMs.
\begin{table}
	\caption{Parameters (name; effective vertical geomagnetic cutoff rigidity calculated for the International Geomagnetic
 Reference Field (IGRF) for the epoch 2010, $P_{\rm c}$; and altitude) of the used NMs along
 with the found scaling factors $\kappa$ (see text) for the four yield functions studied here, Mi13, Ma16, CM12 and CD00.
 Error bars represent the standard error of the mean $\kappa$ over all BRs.
}
	\centering
	\begin{tabular}{c c c | c c c c}
		\hline
NM   & $P_{\rm c}$ (GV) & Alt (m) & $\kappa$ Mi13 &  $\kappa$ Ma16 & $\kappa$ CM12 & $\kappa$ CD00\\
  \hline
Inuvik & 0.3 & 21 & 1.333$\pm$0.003 & 1.237$\pm$0.011 & 1.533$\pm$0.005 & 1.833$\pm$0.011 \\
Apatity & 0.65 & 177 & 1.47$\pm$0.005 & 1.364$\pm$0.012 & 1.69$\pm$0.006 & 2.021$\pm$0.012 \\
Oulu & 0.8 & 15 & 1.199$\pm$0.003 & 1.113$\pm$0.010 & 1.378$\pm$0.004 & 1.649$\pm$0.011 \\
Newark  & 2.4 & 50 & 1.311$\pm$0.006 & 1.207$\pm$0.011 & 1.505$\pm$0.007 & 1.789$\pm$0.011 \\
Moscow & 2.43 & 200 & 1.366$\pm$0.004 & 1.256$\pm$0.009 & 1.569$\pm$0.005 & 1.862$\pm$0.010 \\
Hermanus & 4.58 & 26 & 1.308$\pm$0.003 & 1.142$\pm$0.005 & 1.471$\pm$0.004 & 1.677$\pm$0.005 \\
Athens & 8.53 & 260 & 1.14$\pm$0.007 & 0.927$\pm$0.007 & 1.247$\pm$0.007 & 1.348$\pm$0.007\\
		\hline		
			\label{Tab:NM}
	\end{tabular}\\
\end{table}

\section{Calibration of neutron monitors}
\label{scale_factor}

A NM is an energy-integrating detector, whose count rate at time $t$ in a location with
 the altitude $h$ and the given geomagnetic cutoff $P_{\rm c}$ can be written as
\begin{equation}
N(t,h) = {1\over \kappa}\sum_i{\int_{P_{\rm c}}^\infty{{J_i(P,t)\cdot Y_i(P,h)\cdot dP}}},
\label{Eq:N}
\end{equation}
where the summation is over different types $i$ of primary cosmic rays (protons, helium, etc), $J_i$ is the spectrum
of these particles in the near-Earth space outside the atmosphere and magnetosphere, $P$ is the particle's rigidity,
$Y_i$ is the yield function (YF), and $\kappa$ is a scaling factor (typically in the range 0.8--1.4) correcting for the ``non-ideality''
(local surrounding, exact electronic setup, efficiency of counters, etc) of each NM.
The scaling factors need to be defined experimentally as was done by \citet{usoskin_gil_17} using PAMELA data for protons for 2006\,--\,2010.

\subsection{Contribution of heavier species of cosmic rays}
\label{Sec:h}
\begin{figure*}
	\centering
	\includegraphics[width=\textwidth]{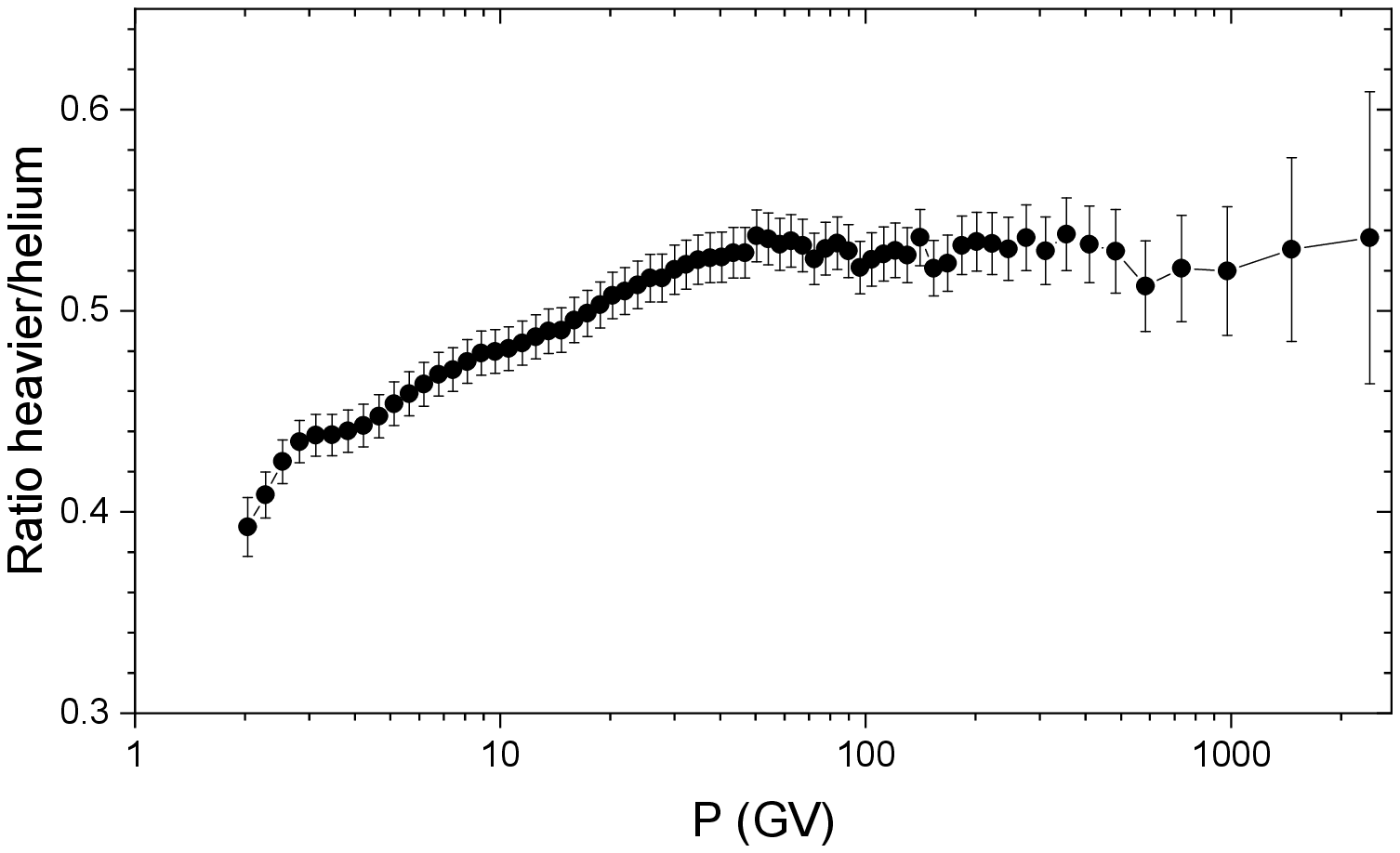}
	\caption{Ratio (in number of helium nuclei, see Equation~\ref{Eq:RR}) of the sum of heavier species to helium as based on the AMS-02 measurements.
    Error bars represent the $1\sigma$ uncertainties of the ratio.}
	\label{Fig:RR}
\end{figure*}

In previous works some assumptions were made. In particular, the nucleonic ratio of the heavier species
 (including all species with $z>$1) to protons in the local interstellar spectrum (LIS) was taken fixed as
 0.3 \citep{webber03,caballero12}, viz. the LIS of heavier species was assumed to proportional to
 the proton LIS.
This assumption could not be directly verified earlier.
Here we can avoid such an \textit{ad-hoc} assumption by using direct measurements of various cosmic-ray species spectra made
 by AMS-02 \citep{aguilar_AMS_17,aguilar_AMS_18_sec,aguilar_AMS_18_N}.
Specifically, we use the measured helium spectra for each BR as an input to the model, assuming that all helium is ${}^{4}$He.
Heavier-than-helium species were not measured for each BR period, but only time-integrated spectra are available
 \citep{aguilar_AMS_17,aguilar_AMS_18_sec,aguilar_AMS_18_N}.
However, since the heliospheric modulation of all heavier species is expected to be similar to that of helium, because
 of the similar charge-to-mass ratio, $Z_i/A_i$, they can be roughly scaled to each other, viz. their ratio in rigidity is expected to be roughly
 constant irrespectively of the modulation level, although this might not be exactly correct.
Therefore, for each heavier specie (Li, Be, B, C, N and O) we assumed that the time-dependent spectrum can be scaled from the
 measured spectrum of helium, e.g., for carbon it can be written as:
\begin{equation}
J_{\rm C}(t) = {\langle J_{\rm C}\rangle \over \langle J_{\rm He}\rangle}\cdot J_{\rm He}(t)
\end{equation}
where $\langle J_{\rm He}\rangle$ and $\langle J_{\rm C}\rangle$ denote helium and carbon spectra, respectively, measured over a long
 period of time between May 2011 and May 2016 (the same stands for other elements).
Although the validity of this assumption cannot be directly validated now, it should be appropriate for the rigidities
  above 3 GV \citep{tomassetti18}, and the contribution of lower-rigidity particles to the NM count rate is less than a few percent
   \citep[see, e.g., Figure 3 in][]{asvestari_JGR_17}.
  The corresponding uncertainty, related to this assumption is small, less than a percent.
Species heavier than oxygen were taken as 1.746 (in nucleonic number) of oxygen \citep[Table 29.1 in][]{tanabashi18}.
Heavier elements with $z>2$ are summed up
 to produce the rigidity-dependent ratio (in number of $\alpha-$particles) of heavier species to helium
\begin{equation}
R(P) = {1\over 4}\sum_i A_i{\langle J_i(P)\rangle\over \langle J_{\rm He}(P)\rangle}
\label{Eq:RR}
\end{equation}
which is shown in Figure~\ref{Fig:RR}.
One can see that the ratio slightly varies with the particle rigidity but on the average it is 0.53, so that the helium spectrum scaled up with
 a factor of 1.53 can roughly represent all heavier-than-proton species of cosmic rays.
Here we used exactly the ratio shown in Figure~\ref{Fig:RR}.

\subsection{NM yield functions}
\label{Sec:YF}
Precise measurements of the cosmic ray spectra, performed for various modulation conditions outside the Earth's atmosphere, make it possible
 to validate NM yield functions (see Equation~\ref{Eq:N}).
Here we tested several modern yield functions which are widely used in the community.

\textit{Mi13} yield function was computed by \citet{mishev13} (including the erratum) as a result of a full Monte-Carlo simulation of the
 cosmic-ray induced atmospheric cascade, explicitly considering the finite lateral extent of the cascade and the detector's dead time.
We used the following approximation of the Mi13 yield function per nucleon, $Y$, (in units m$^2$ sr, for a 6NM64 neutron monitor at the sea-level):
\begin{equation}
\ln(Y) = a\cdot X^3 +b\cdot X^2 + c\cdot X + d,	
\label{Eq:YF}
\end{equation}
where $X = 0.5\cdot\ln(T^2+1.876\cdot T)$ and $T$ is the kinetic energy in GeV per nucleon.
Coefficients $a$ through $d$ are given in Table~\ref{Tab:YF} for both protons and helium.

\textit{Ma16} yield function was computed by \citet{mangeard2} using a full Monte-Carlo simulation of the cascade.
Parameterization was also made using Equation~\ref{Eq:YF} with the coefficients listed in Table~\ref{Tab:YF}.

\textit{CM12} yield function was obtained by \citet{caballero12} in a semi-empirical way using a latitude survey of a standard NM.
Here we used the parameterization as provided by \citet{maurin15}.

\textit{CD00} was computed by \citet{clem00} using Monte-Carlo, and is still in use in the research community.
Here we used the parameterization as provided by \citet{maurin15}.

\begin{table}
\caption{Coefficients for parameterization of Mi13 and Ma16 yield functions (Equation~\ref{Eq:YF}).}
\begin{tabular}{c|cccc}
$T$ range & $a$ & $b$ & $c$ & $d$ \\
\hline
&\multicolumn{4}{c}{Mi13 protons}\\
$<1.25$ GeV & 6.09 & -14.06 & 13.98 & -11.615\\	
1.25\,--\,10 GeV & -0.186 & 0.428 & 2.831 & -8.76\\
$>10$ GeV & 0 & -0.0365 & 1.206 &-4.763\\	
\hline
&\multicolumn{4}{c}{Mi13 helium}\\
$<1.6$ GeV/nuc & 2.0404 & -8.1776 & 12.354 & -11.1\\	
1.6\,--\,15 GeV/nuc & 0.1179 & -1.2022 & 4.9329 & -8.65\\
$>15$ GeV/nuc & 0 & -0.0365 & 1.206 &-4.763\\
\hline
&\multicolumn{4}{c}{Ma16 protons}\\
$<2.5$ GeV & 1.8289 & -4.7471 & 7.3371 & -9.5333\\	
2.5\,--\,14 GeV & 0 & -0.4484 & 3.3632 & -7.8\\
$>14$ GeV & 0 & 0 & 0.9284 &-4.473\\
\hline
&\multicolumn{4}{c}{Ma16 helium}\\
$<1.8$ GeV/nuc & 1.3797 & -3.9957 & 6.1898 & -8.232\\	
1.8\,--\,14 GeV/nuc & 0.2901 & -2.0801 & 6.02 & -8.8583\\
$>14$ GeV/nuc & 0 & 0 & 0.9304 &-4.557\\
\hline
\end{tabular}
\label{Tab:YF}
\end{table}

\begin{figure*}
	\centering
	\includegraphics[width=\textwidth]{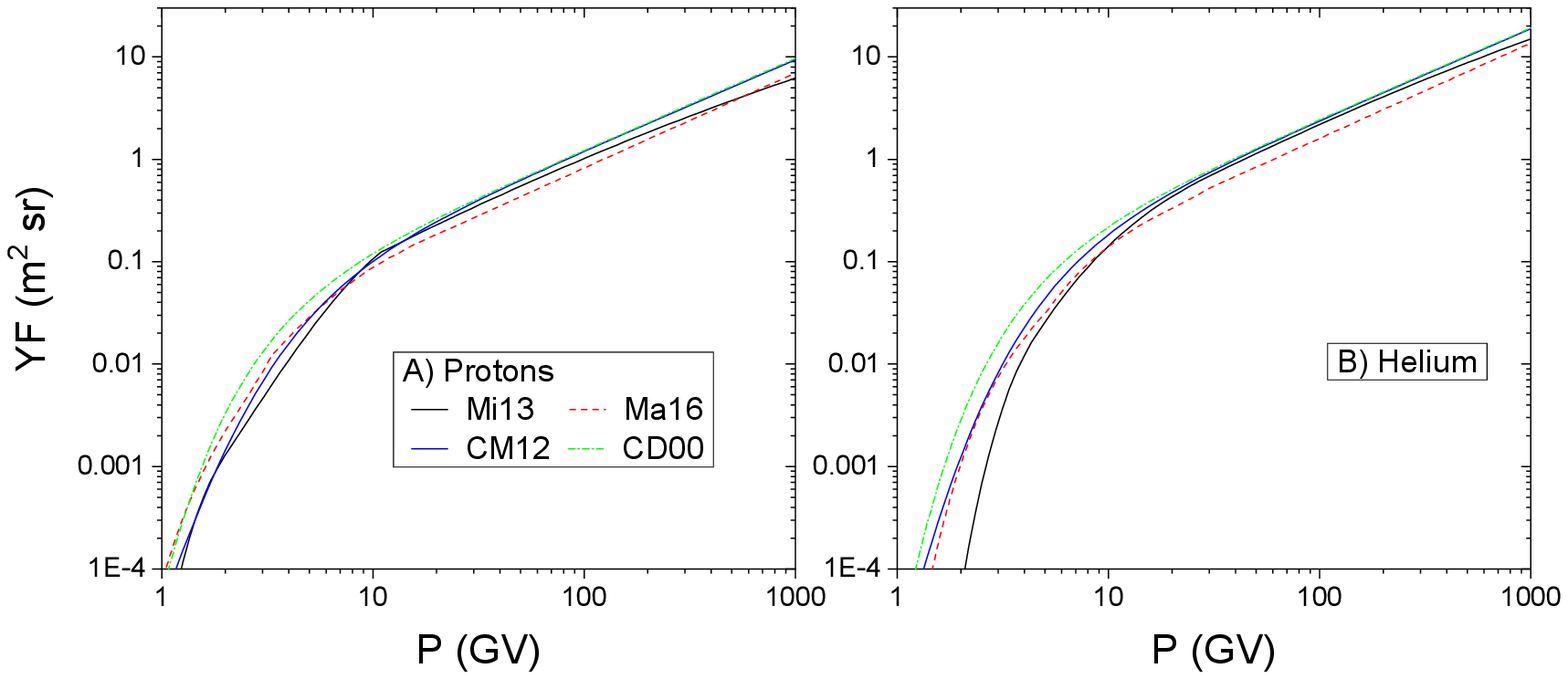}
	\caption{Yield functions of a standard sea-level 6NM64 neutron monitor for primary protons (panel A) and helium (per particle, panel B)
        for the four models used here.  }
	\label{Fig:YF}
\end{figure*}

\begin{figure}
	\centering
	\includegraphics[width=\columnwidth]{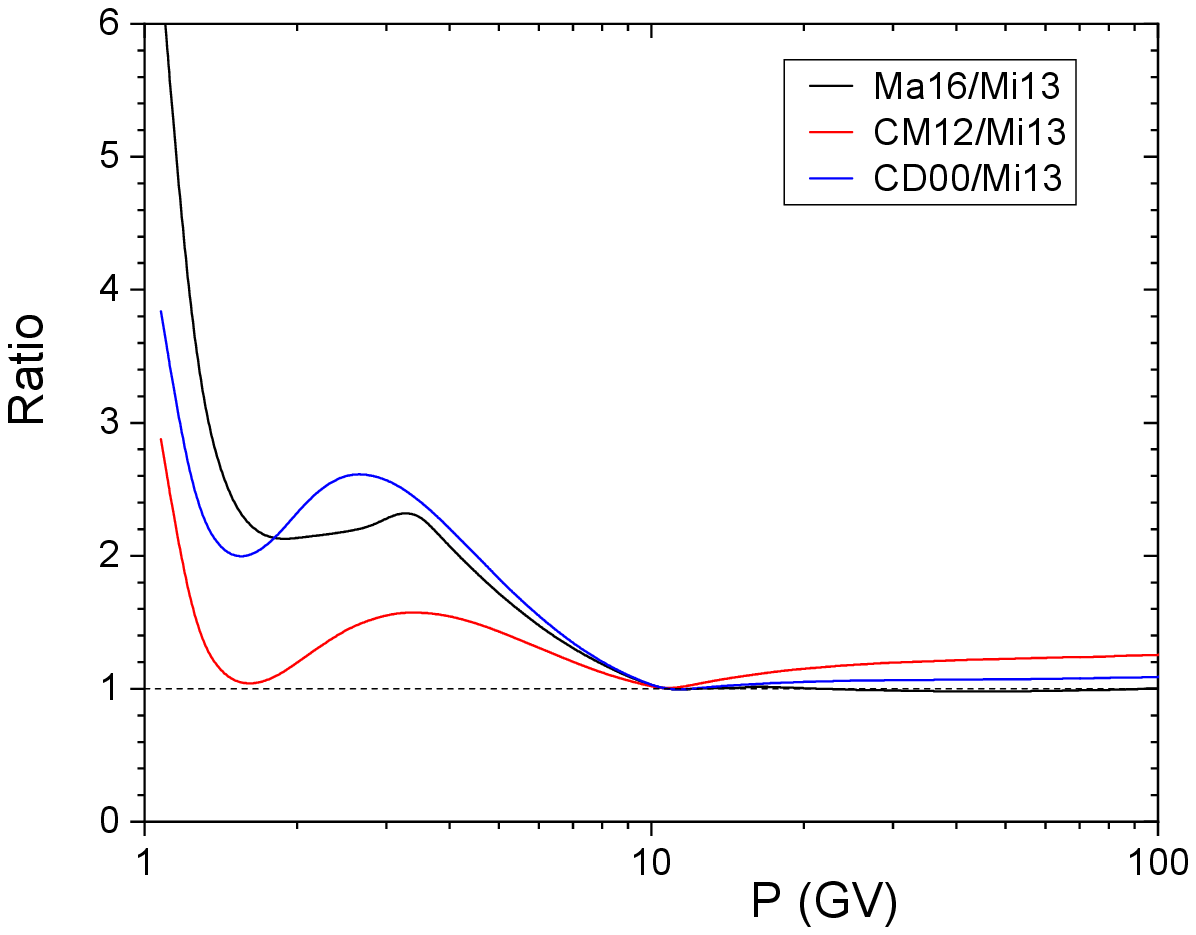}
	\caption{Normalized ratios of the Ma16, CM12 and CD00 yield functions for protons to that by Mi13 as a function
        of particles rigidity.
    All yield functions are normalized to unity at 10 GV rigidity.}
	\label{Fig:YF_R}
\end{figure}

The yield functions are shown in Figure~\ref{Fig:YF}.
While they have similarly looking shapes, they differ in details by up to a factor of two.
The normalized ratio of the three yield functions to the Mi13 one is shown in Figure~\ref{Fig:YF_R}.
While all of them agree pretty well in the high-rigidity tail above 10 GV, the difference is significant
 in the lower-rigidity range of several GV, which may affect the modelled count rates of NMs.

Here we performed a full re-calibration of the NM response and verified the validity of the yield functions using, as $J_i(P,t)$, the spectra of protons and
 helium (the latter scaled to account for heavier species, as described above) directly measured by AMS-02.
We computed, using Equation~\ref{Eq:N}, the expected count rates of seven stable NMs (see Table~\ref{Tab:NM}) covering the range from low to high
 geomagnetic latitudes, for each BR between 2426\,--\,2506, directly using the AMS-02 data as the GCR spectrum.
Next, we compared these expected count rates with the measured ones for each BR, and calculated the relevant scaling factor $\kappa$ (Eq.~\ref{Eq:N}),
 defined as the ratio of the modelled to measured count rates.
The mean values of the scaling factors are shown in Table~\ref{Tab:NM}.
For the Mi13 yield function they agree within several percent with those based on PAMELA data \citep{usoskin_gil_17}.
The scaling factors are typically somewhat greater than unity, indicating that the actual efficiency of a NM is slightly lower than that of an ideal NM.
This is related to the local environment (e.g., constructions above the NM) or electronic setup (high voltage or dead time).
It is expected that a NM with the standard dead time of about 20 $\mu$sec has a slightly lower count rate than an ideal NM with the full multiplicity counts.
The systematically higher values of $\kappa$ for Apatity and Moscow NMs can be explained by the slightly ($\approx$15\%) lower efficiency
 of Soviet analogs (SNM-15) with respect to the original NM64 (produced by the Chalk River laboratory) counters \citep{abunin11,gil15}.
On the other hand, CD00 yield function leads to too high scaling factors, suggesting that it may overestimate the NM response.

\begin{figure}
	\centering
	\includegraphics[width=\columnwidth]{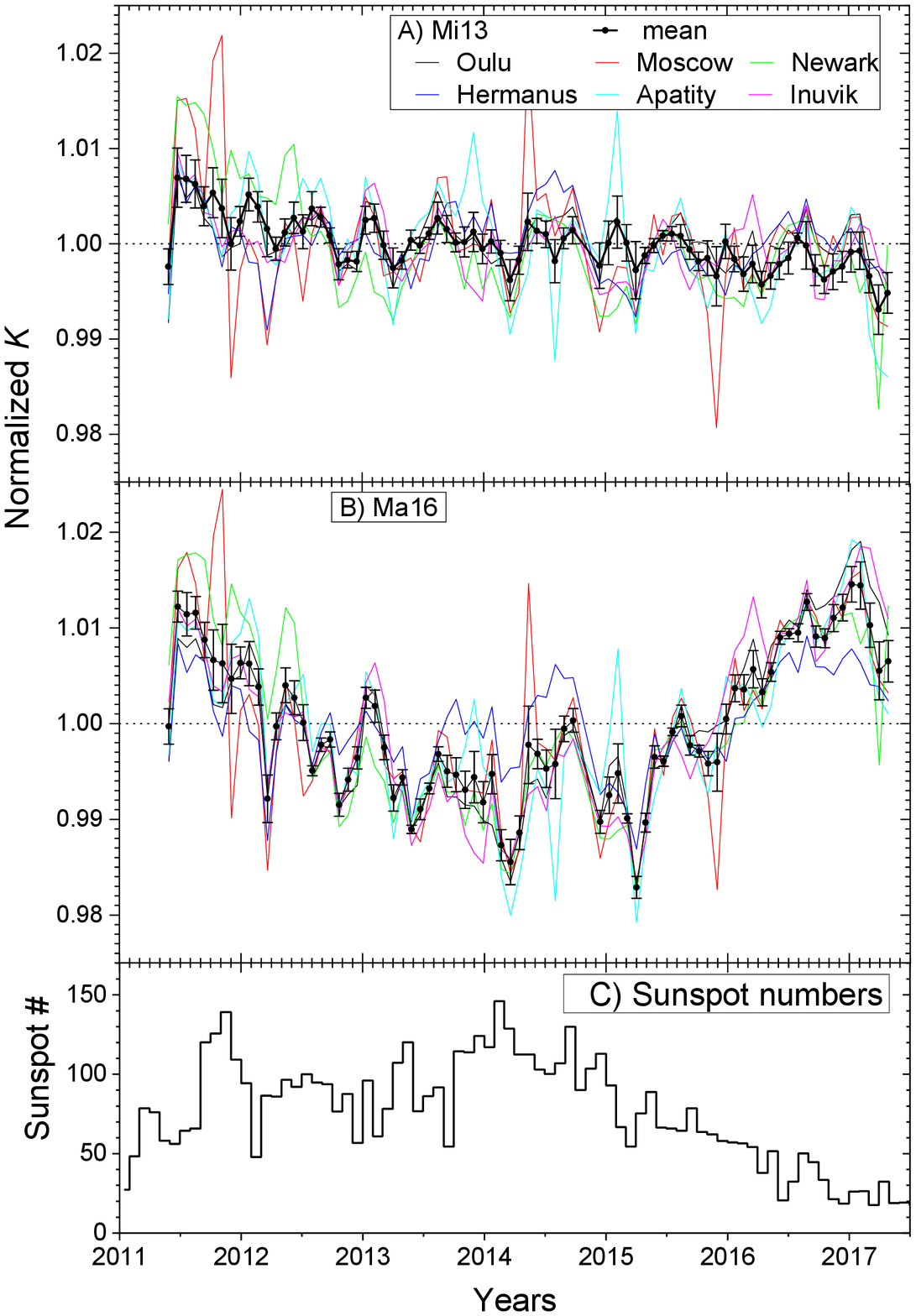}
	\caption{Scaling factors $\kappa$, normalized to the mean values during the covered time period, for different neutron monitors
     (color curves) and their mean (black) for Mi13 (panel A)
       and Ma16 (panel B) yield function. The international sunspot number (v.2.0, http://www.sidc.be/silso/datafiles) is shown as reference in panel C.}
	\label{Fig:kappa}
\end{figure}

\begin{figure*}
	\centering
	\includegraphics[width=\textwidth]{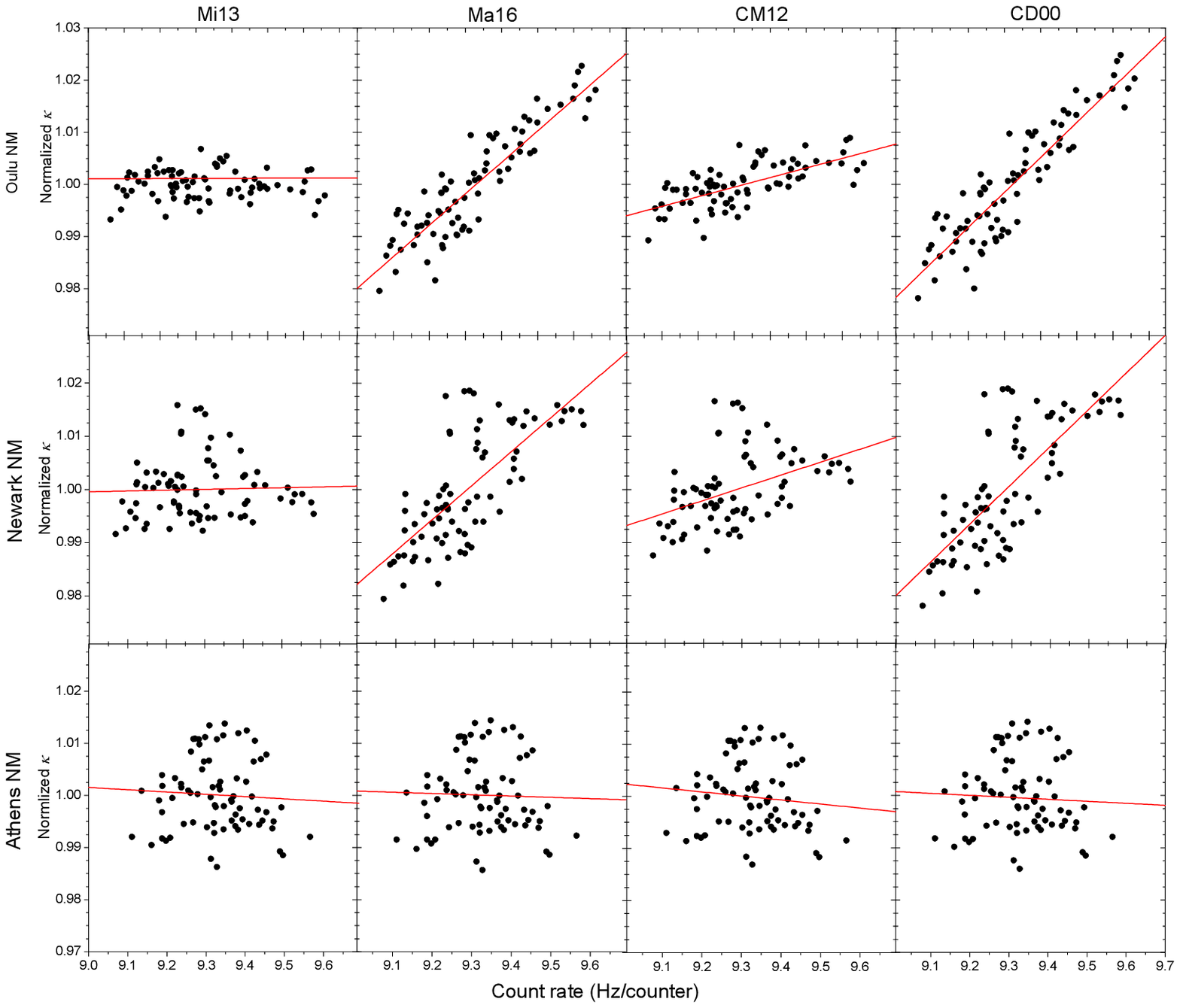}
	\caption{Scatter plots of the normalized scaling factor $\kappa$ vs. the count rates of three NMs (Oulu, Newark and Athens) for individual
      BRs, for the four yield functions analyzed here.
      Red lines correspond to the best-fit linear trends. }
	\label{Fig:res}
\end{figure*}

\begin{figure}
	\centering
	\includegraphics[width=\columnwidth]{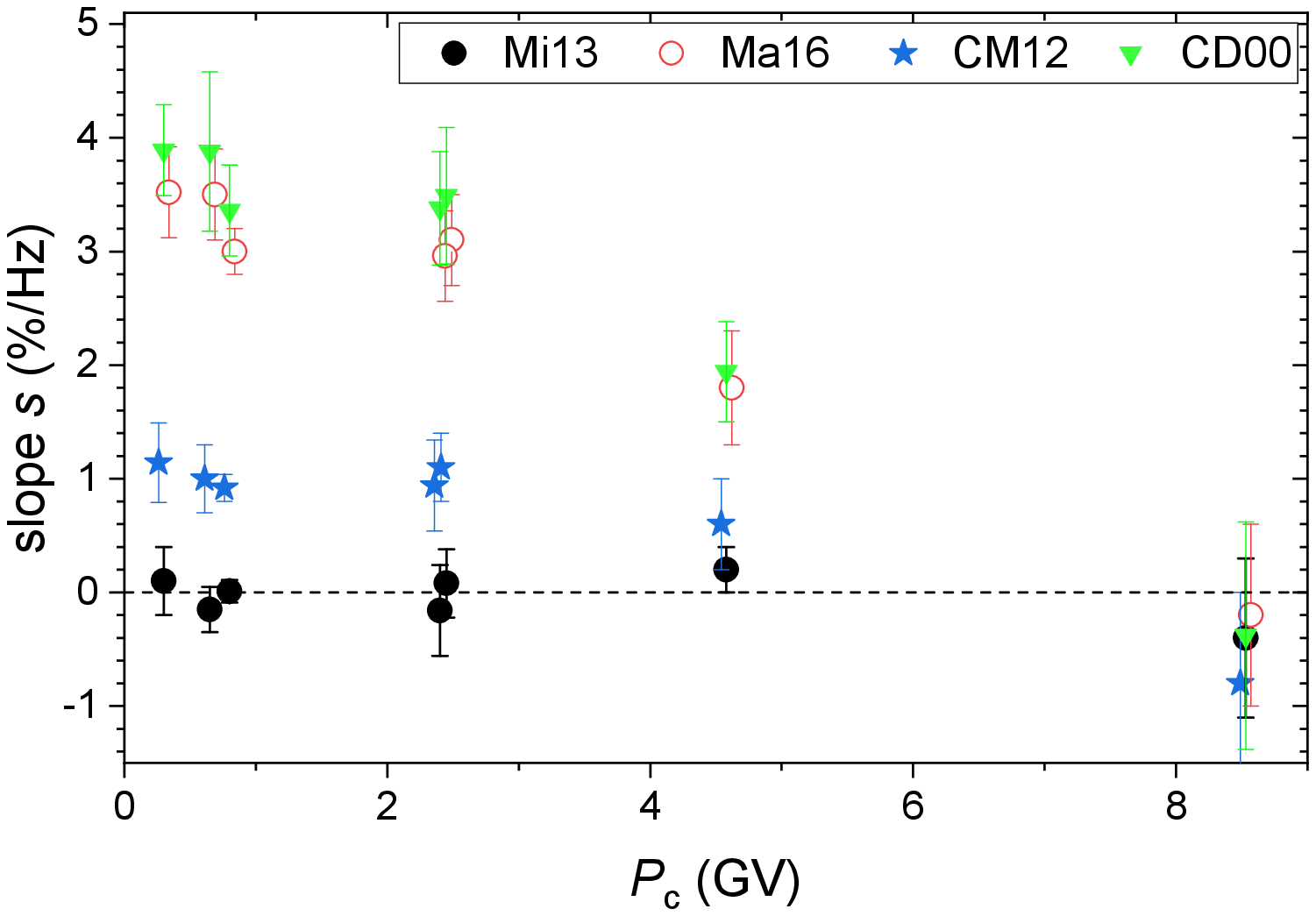}
	\caption{The slope of the linear regression between the scaling factor $\kappa$ and the count rate (see Table~\ref{Tab:slope}), along with
     the standard error, for different yield functions (colors as denoted in the legend) and NMs, characterized by their geomagnetic cutoff
     rigidity $P_{\rm c}$.}
	\label{Fig:slope}
\end{figure}

Since the new AMS-02 data cover a wide range of solar modulation, from low activity in 2011 to the maximum in 2014 and again to the declining
 phase towards a new minimum (see Figure~\ref{Fig:kappa}C), in addition to computing the mean scaling factors we can also check their stability in time.
If the model works correctly, the value of $\kappa$ should not depend on the exact level of solar activity.
On the contrary, if it does, the yield function is likely incorrect.
The obtained values of $\kappa$, normalized to the averaged value during the analyzed time period, are shown in Figure~\ref{Fig:kappa}
 for Mi13 and Ma16 models, for different NMs.
One can see that the $\kappa-$factor is stable, within $\pm 0.5$\% without any visible trend, for the Mi13 yield function (panel A).
In contrast, the $\kappa$-values calculated for the Ma16 yield function (Fig.~\ref{Fig:kappa}B) exhibit a clear wave with the magnitude of $\approx 3$\%,
 which corresponds to the 11-year solar cycle.
In order to analyse this in more details, we have produced scatter plots of the $\kappa-$factors calculated using
 different yield functions vs. the count rates of each NM for individual BR.
Some of such plots are shown in Figure~\ref{Fig:res}.
One can see that in some cases there is a clear and strong positive correlation between the two quantities (e.g., Oulu NM for Ma16), but in other cases they
 are independent of each other (e.g., Athens NM for all yield functions).
In order to quantify that, each of such scatter plot was analyzed and the corresponding Pearson's correlation coefficient $r$ and the linear regression
 slope $s$ were calculated and collected in Table~\ref{Tab:slope}.
It is important that there is no statistically significant relation between $\kappa$ and count rates for all NMs when the Mi13 yield function
 is used (see first column in Table~\ref{Tab:slope}).
On the contrary, for all other yield functions the correlation is significant for most of the NMs, except for Athens NM with the geomagnetic cutoff rigidity
 of 8.53 GV.
In order to illustrate this, we show in Figure~\ref{Fig:slope} the regression slopes $s$ as function of the geomagnetic
 cutoff rigidities $P_{\rm c}$ for the four yield functions.
One can see that the Mi13 produces no significant relation even for the high-latitude NMs.
On the other hand, all other yield functions lead to a statistically significant relation between the two quantities
 for polar NMs which fades down towards higher cutoff rigidities and disappears for $P_{\rm c}=8.53$ GV.
The strongest relation (as high as 3.5\,--\,4 \%/Hz) is for Ma16 and CD00 yield functions and quite moderate ($\approx$1 \%/Hz)
 for CM12 one.
This implies that these yield functions may introduce an additional bias of up to 10\% within a solar cycle, where the
 NM count rate varies by 1\,--\,2 Hz/counter.

This implies that most of the yield functions tend to \textit{overestimate} the response of a NM to low-energy
 cosmic rays (below several GV).
This leads to a positive relation between $\kappa$ and the count rate.
On the other hand, the fact, that the relation is absent even for polar NMs for the Mi13 yield function, suggests that the
 low-energy part of this yield function is correct.
This is observed in Figure~\ref{Fig:YF_R} as an excess of all yield functions over the Mi13 one in the rigidity range below 10 GV.
Although this effect is not large for GCRs, within 10\%, it may be important for a study of much softer solar energetic particles,
 leading to a potential underestimate of their flux.
Since all the yield functions lead to no relation for Athens NM, we may safely assume that all of them correctly
 reproduce the NM response in the rigidity range above 8 GV.
\begin{table}
\caption{Relation between the scaling factors $\kappa$ and count rates for the seven NMs and four yield functions studied here:
  the slope $s$ (in \%/Hz -- see text) and the Pearson correlation coefficient $r$.
  The values consistent with the null hypothesis of no relation are highlighted in bold.}
\begin{tabular}{lc|cccc}
Name	&		&		Mi13		&		Ma16		&		CM12	& CD00	\\
\hline
Inuvik & $s$  & \textbf{0.1 $\pm$ 0.3} & 3.5 $\pm$ 0.4 & 1.14 $\pm$ 0.4 & 3.89 $\pm$ 0.4\\
 & $r$ & \textbf{0.08 $\pm$ 0.11} & 0.9 $\pm$ 0.02 & 0.66 $\pm$ 0.06 & 0.90 $\pm$ 0.02 \\
\hline
Apatity & $s$  & \textbf{-0.15 $\pm$ 0.2} & 3.5 $\pm$ 0.4 & 1 $\pm$ 0.3 & 3.88 $\pm$ 0.7\\
 & $r$ & \textbf{-0.07 $\pm$ 0.11} & 0.74 $\pm$ 0.04 & 0.37 $\pm$ 0.09 & 0.76 $\pm$ 0.04 \\
\hline
Oulu & $s$  & \textbf{0.01 $\pm$ 0.1} & 3 $\pm$ 0.2 & 0.92 $\pm$ 0.12 & 3.36 $\pm$ 0.4 \\
 & $r$ & \textbf{-0.01 $\pm$ 0.11} & 0.9 $\pm$ 0.03 & 0.67 $\pm$ 0.05 & 0.91 $\pm$ 0.02\\
\hline
Moscow & $s$  & \textbf{-0.16 $\pm$ 0.4} & 2.96 $\pm$ 0.4 & 0.94 $\pm$ 0.4 & 3.38 $\pm$ 0.5\\
 & $r$ & \textbf{-0.09 $\pm$ 0.12} & 0.83 $\pm$ 0.04 & 0.49 $\pm$ 0.09 & 0.85 $\pm$ 0.03 \\
\hline
Newark & $s$  & \textbf{0.08 $\pm$ 0.3} & 3.1 $\pm$ 0.4 & 1.1 $\pm$ 0.3 & 3.49 $\pm$ 0.6 \\
 & $r$ & \textbf{0.03 $\pm$ 0.11} & 0.71 $\pm$ 0.05 & 0.4 $\pm$ 0.1 & 0.74 $\pm$ 0.05\\
\hline
Hermanus & $s$  & \textbf{0.2 $\pm$ 0.2} & 1.8 $\pm$ 0.5 & 0.6 $\pm$ 0.4 & 1.94 $\pm$ 0.44\\
 & $r$ & \textbf{0.09 $\pm$ 0.11} & 0.68 $\pm$ 0.05 & 0.32 $\pm$ 0.1 & 0.71 $\pm$ 0.06\\
\hline
Athens & $s$  & \textbf{-0.4$\pm$0.7} & \textbf{-0.2 $\pm$ 0.8} & \textbf{-0.8 $\pm$ 0.8} & \textbf{-0.38 $\pm$ 1.0}\\
 & $r$ & \textbf{-0.06 $\pm$ 0.1} & \textbf{-0.03 $\pm$ 0.11} & \textbf{-0.11 $\pm$ 0.11} & \textbf{-0.05 $\pm$ 0.11} \\
\hline
\end{tabular}
\label{Tab:slope}
\end{table}

It is interesting to note that data of Oulu NM are closer to the model expectations than other analyzed NMs, as observed
 by the least spread of points in Figure~\ref{Fig:res} and the smallest uncertainty in the slope/regressions values (Table~\ref{Tab:slope})
 among all NMs.
This is in line with the earlier drawn conclusions that Oulu NM is one of the most stable NMs used as a reference one
 \citep[e.g.][]{ahluwalia13,usoskin_gil_17}.

%
%

\section{Validation of the force-field approximation}
\label{phi}

\subsection{Fitting the data}

In this section we describe the work performed to check the validity of the force-field approximation.
Spectra of protons measured for each BR were individually fitted with the force-field model to
 define the corresponding modulation potential $\phi$.
The force-field model links the energy spectrum of GCR particles of a given type $i$ (protons, helium, etc.)
 near Earth, $J$, with their reference intensity outside the heliosphere, LIS
 $J_{\rm LIS}$, so that
\begin{linenomath*}
	\begin{equation}\label{Eq:FF}
	J_{i}(T,\phi)=J_{{\rm LIS}_{i}}(T+\Phi_{i})\frac{T(T+2T_{\rm r})}{(T+\Phi_{i})(T+\Phi_i+2T_{\rm r})}
	\end{equation}
\end{linenomath*}
where $T$ is the kinetic energy per nucleon, $T_{\rm r}=0.938$ GeV is the proton's rest mass, and
 $\Phi_i=\phi\cdot(eZ_{i}/A_{i})$.
The force-field approximation is obtained as an analytical solution (in the form of characteristic curves) of the heavily
 simplified GCR transport equation \citep{gleeson68,caballero04}, where all the modulation effects are reduced to a single
 parameter $\phi$ called the modulation potential.
Although it has little physical sense because of the simplified assumptions (spherical symmetry, steady state, adiabatic changes),
 the force-field model provides a very good and useful parameterization of the near-Earth GCR spectrum \citep[e.g.][]{vainio09}.
The exact value of the modulation parameter depends on the reference LIS \citep{usoskin_Phi_05,herbst10,asvestari_JGR_17}.
 Here we used a recent estimate of the proton LIS by \citet{vos15}, parameterized as:
\begin{equation}
J_{{\rm LIS}_{\rm p}} = 2.7\cdot 10^3\, \,{T^{1.12}\over \beta^2}\left({T+0.67 \over 1.67}\right)^{-3.93},
\label{Eq:LIS}
\end{equation}
where $\beta=v/c$ is the ratio of the proton's velocity to the speed of light, $J$ and $T$ are given in units of
[m$^2$ sec sr GeV/nuc]$^{-1}$ and GeV/nuc, respectively.

The measured proton spectra were fitted with the force-field model (Equations~\ref{Eq:FF} and \ref{Eq:LIS}) using
 the $\chi^2$ method in the same way as described by \citet{koldobsky18}.
The merit function $\chi^2$ was calculated as
\begin{equation}
\label{Eq:chi2}
\chi^2 = \sum_j\left({J_{{\rm mod},j} - J_{{\rm meas},j}\over \sigma_j}\right)^2,
\end{equation}
where summation is over the rigidity bins in the measured spectrum, $J_{\rm meas}$,
 and $\sigma_j$ is the uncertainty of the measured intensity,
 while the modelled intensity, $J_{{\rm mod},j}$ is an integral of the
 force-field spectrum (Equation~\ref{Eq:FF}) inside the $j-$th rigidity bin in the AMS-02 data computed as
\begin{equation}
J_{{\rm mod},j}={1\over {P_{j,2}-P_{j,1}}}\int_{P_{j,1}}^{P_{j,2}}{J_{\rm mod}(P')\,dP'},
\end{equation}
where $P_{j,1}$ and $P_{j,2}$ are the lower and upper bounds of the $j-$th rigidity bin in the AMS-02 data.
The best-fit value of $\phi$ was defined as the one minimizing the merit function to $\chi^2_{\rm min}$, and its $\pm\sigma$ interval
 defined as corresponding to the $\chi^2_{\rm min}+1$.

\begin{figure}[t]
	\centering
	\includegraphics[width=\columnwidth]{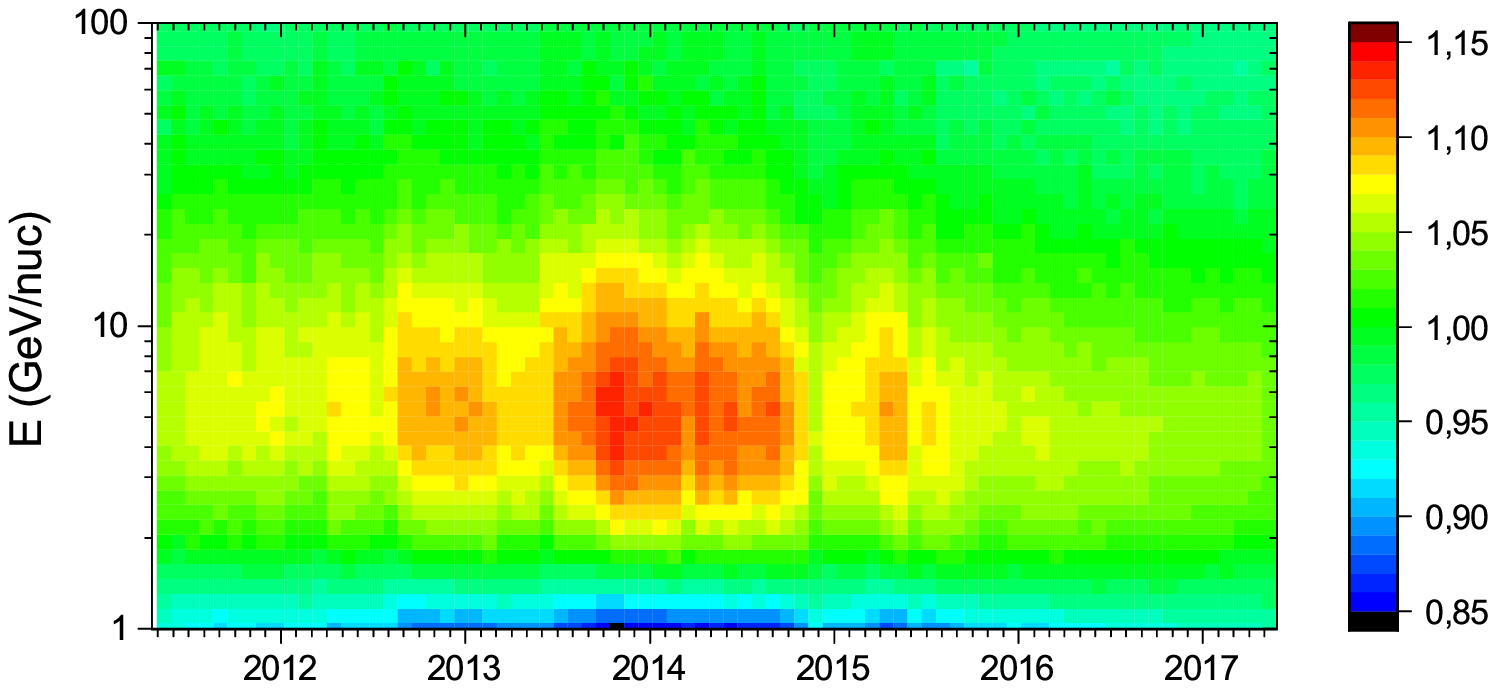}
	\caption{Ratio of the best-fit force-field model \citep[LIS by][]{vos15} to measured proton AMS-02 spectra.}
	\label{Fig:p_fit}
\end{figure}

\begin{figure*}
	\centering
	\includegraphics[width=\textwidth]{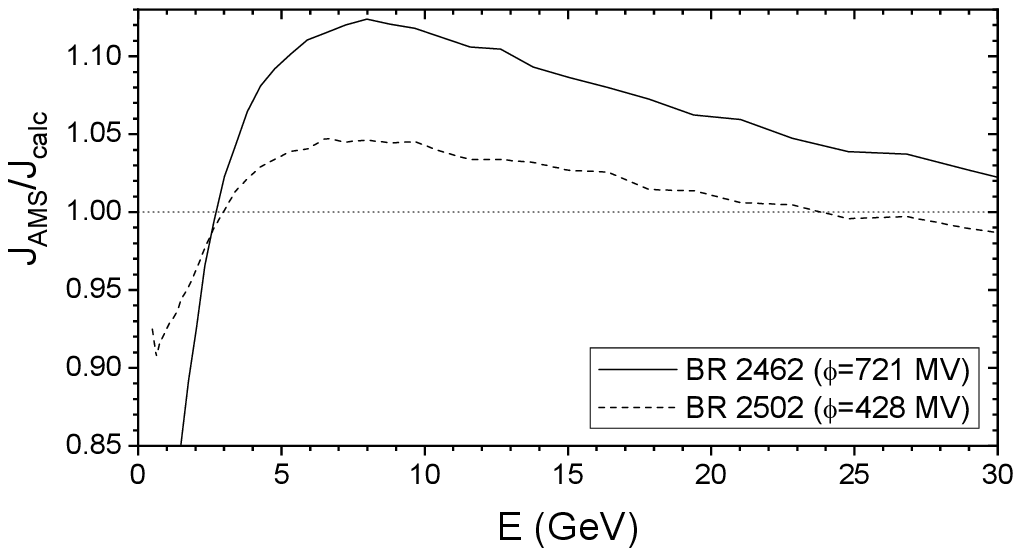}
	\caption{Ratio of the best-fit to measured proton spectra for two BRs, 2462 and 2502,
 corresponding to solar-cycle maximum and minimum, respectively. }
	\label{Fig:p_example_fit}
\end{figure*}

Here we fitted the proton spectra in the energy range 1\,--\,30 GeV/nucleon, which is affected by solar modulation.
This includes $n=31$ energy bins (30 degrees of freedom).
The ratio of the best-fit to the measured spectra is shown in Figure~\ref{Fig:p_fit} as function of energy (Y-axis) and time (X-axis).
One can see that the fit agrees with the data within $\pm 10$\% for the solar maximum period and within a few \% during the
 ascending and descending phases of the solar cycle.
Two cuts, corresponding to high (BR 2462) and low (BR 2502) modulation conditions are shown in Figure~\ref{Fig:p_example_fit}.
During quiet periods, the quality of the fit is better than for periods with high activity.
This is understandable, since solar modulation is more complicated, in particular by propagating barriers, during active
 than during quiet periods.
Nevertheless, the fits are good  even during active periods.
We note that the force-field approximation is not thought to precisely reproduce the exact spectrum of particles near Earth
 \citep[e.g.,][]{geiseler18,mangeard18}, but only provides a reasonable parameterization for it.
The obtained 10\% accuracy is fairly good for practical purposes of assessing the cosmic-ray-related atmospheric effects
 and fully validates the use of the force-field model for applications, while it should not be applied for a detailed study of cosmic rays.

\begin{figure}
	\centering
	\includegraphics[width=\columnwidth]{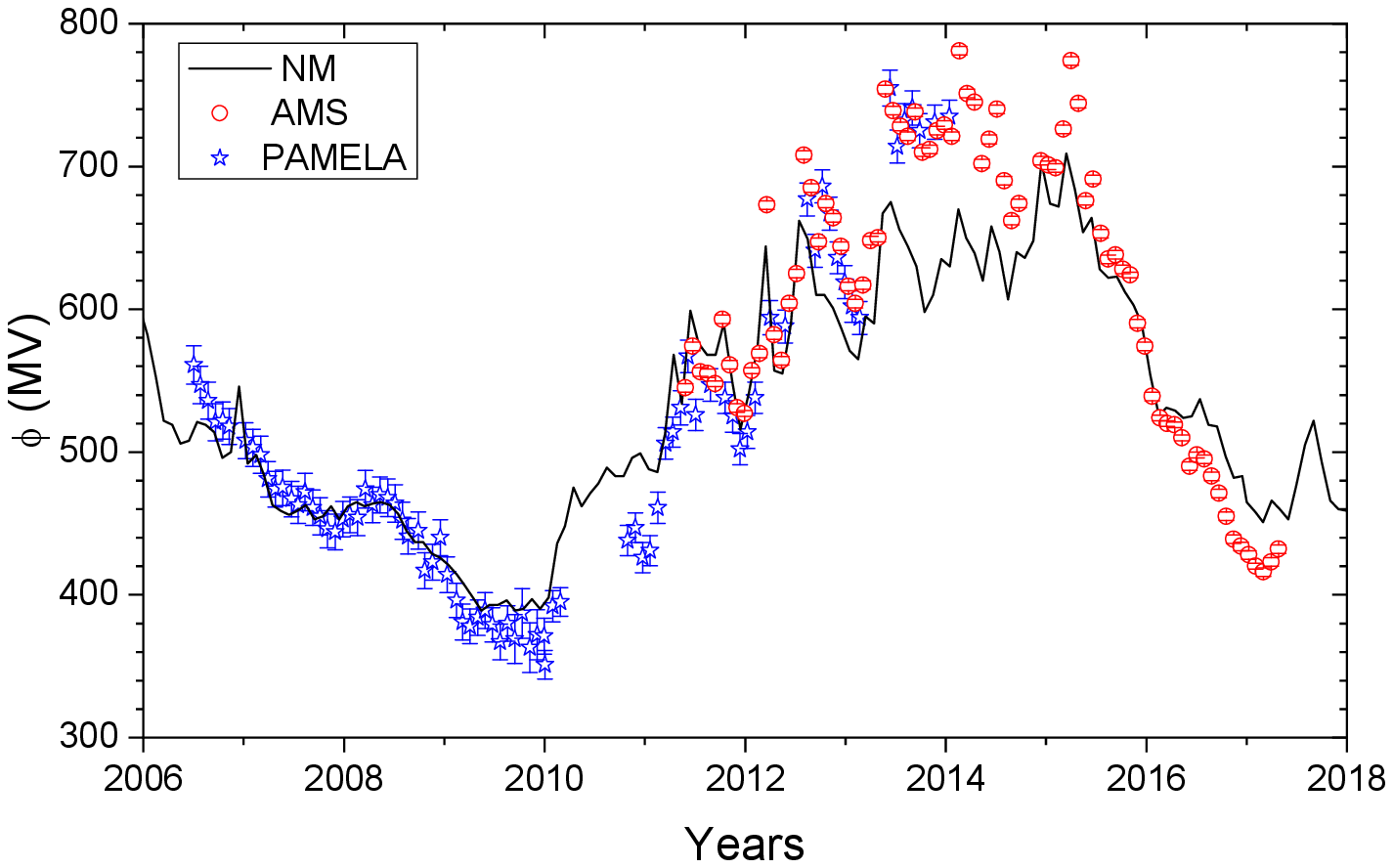}
	\caption{Values of the modulation potential $\phi$ reconstructed here from space-borne (symbols) and
    NM-based (solid curve) data.
  AMS-based values were obtained here (see also Table~\ref{Tab:phi}), PAMELA and NM-based values were adopted from \citet{koldobsky18}.}
	\label{Fig:time}
\end{figure}

Figure~\ref{Fig:time} shows the time dependence of $\phi$-value obtained above from the AMS-02 data (see Table~\ref{Tab:phi}) along with
 those estimated from PAMELA data by \citet{koldobsky18} and reconstructed values from the NM network \citep[][see also http://cosmicrays.oulu.fi/phi/phi.html]{usoskin_gil_17}.
We note that the NM-based $\phi-$values were normalized to PAMELA data for 2006\,--\,2010, and thus the new AMS-02
 data serves as a direct test for the reconstruction.
One can see that the AMS-based values lie close to the PAMELA-based ones during the time of their overlap.
On the other hand, the NM-based modulation potential looks systematically too low, by 50\,--\,70 MV during the solar maximum,
 and agrees with the AMS-based ones for the times of moderate solar activity and then exceeds the fitted values by $\approx 40$ MV after 2016.
Overall, the NM-based values of $\phi$ agrees with the direct fits to AMS-02 data within 50 MV.

\begin{table}
\caption{The values of the modulation potential $\phi$ (in MV) obtained by a fit to the proton
 spectra measured by AMS-02 for each BR.
 The best-fit value and its standard error are given.}
\begin{tabular}{cc|cc|cc|cc}
\hline
BR & $\phi$ (MV) & BR & $\phi$ (MV) & BR & $\phi$ (MV) & BR & $\phi$ (MV) \\
\hline
2426 & 545$\pm$3 & 2446 & 664$\pm$3 & 2466 & 702$\pm$3 & 2488 & 574$\pm$3 \\
2427 & 574$\pm$3 & 2447 & 644$\pm$3 & 2467 & 719$\pm$3 & 2489 & 539$\pm$3 \\
2428 & 556$\pm$3 & 2448 & 616$\pm$3 & 2468 & 740$\pm$3 & 2490 & 524$\pm$2 \\
2429 & 555$\pm$3 & 2449 & 604$\pm$3 & 2469 & 690$\pm$3 & 2491 & 520$\pm$2 \\
2430 & 548$\pm$2 & 2450 & 617$\pm$3 & 2470 & 662$\pm$3 & 2492 & 519$\pm$2 \\
2431 & 593$\pm$3 & 2451 & 648$\pm$3 & 2471 & 674$\pm$3 & 2493 & 510$\pm$2 \\
2432 & 561$\pm$3 & 2452 & 650$\pm$3 & 2474 & 704$\pm$3 & 2494 & 490$\pm$3 \\
2433 & 531$\pm$3 & 2453 & 754$\pm$3 & 2475 & 701$\pm$3 & 2495 & 498$\pm$2 \\
2434 & 527$\pm$2 & 2454 & 739$\pm$3 & 2476 & 699$\pm$3 & 2496 & 495$\pm$3 \\
2435 & 557$\pm$2 & 2455 & 728$\pm$3 & 2477 & 726$\pm$3 & 2497 & 483$\pm$3 \\
2436 & 569$\pm$3 & 2456 & 721$\pm$3 & 2478 & 774$\pm$3 & 2498 & 471$\pm$3 \\
2437 & 673$\pm$3 & 2457 & 738$\pm$3 & 2479 & 744$\pm$3 & 2499 & 455$\pm$3 \\
2438 & 582$\pm$3 & 2458 & 710$\pm$3 & 2480 & 676$\pm$3 & 2500 & 439$\pm$3 \\
2439 & 564$\pm$3 & 2459 & 712$\pm$3 & 2481 & 691$\pm$3 & 2501 & 434$\pm$3 \\
2440 & 604$\pm$3 & 2460 & 725$\pm$3 & 2482 & 653$\pm$3 & 2502 & 428$\pm$3 \\
2441 & 625$\pm$3 & 2461 & 729$\pm$3 & 2483 & 635$\pm$3 & 2503 & 420$\pm$2 \\
2442 & 708$\pm$3 & 2462 & 721$\pm$3 & 2484 & 638$\pm$3 & 2504 & 416$\pm$3 \\
2443 & 685$\pm$3 & 2463 & 781$\pm$3 & 2485 & 628$\pm$3 & 2505 & 423$\pm$3 \\
2444 & 647$\pm$3 & 2464 & 751$\pm$3 & 2486 & 624$\pm$2 & 2506 & 432$\pm$3 \\
2445 & 674$\pm$3 & 2465 & 745$\pm$2 & 2487 & 590$\pm$3 &  &    \\
\hline
\end{tabular}
\label{Tab:phi}
\end{table}

\subsection{Reconstruction of the modulation potential from NM}
\label{Sec:rec}
\begin{figure}
	\centering
	\includegraphics[width=\columnwidth]{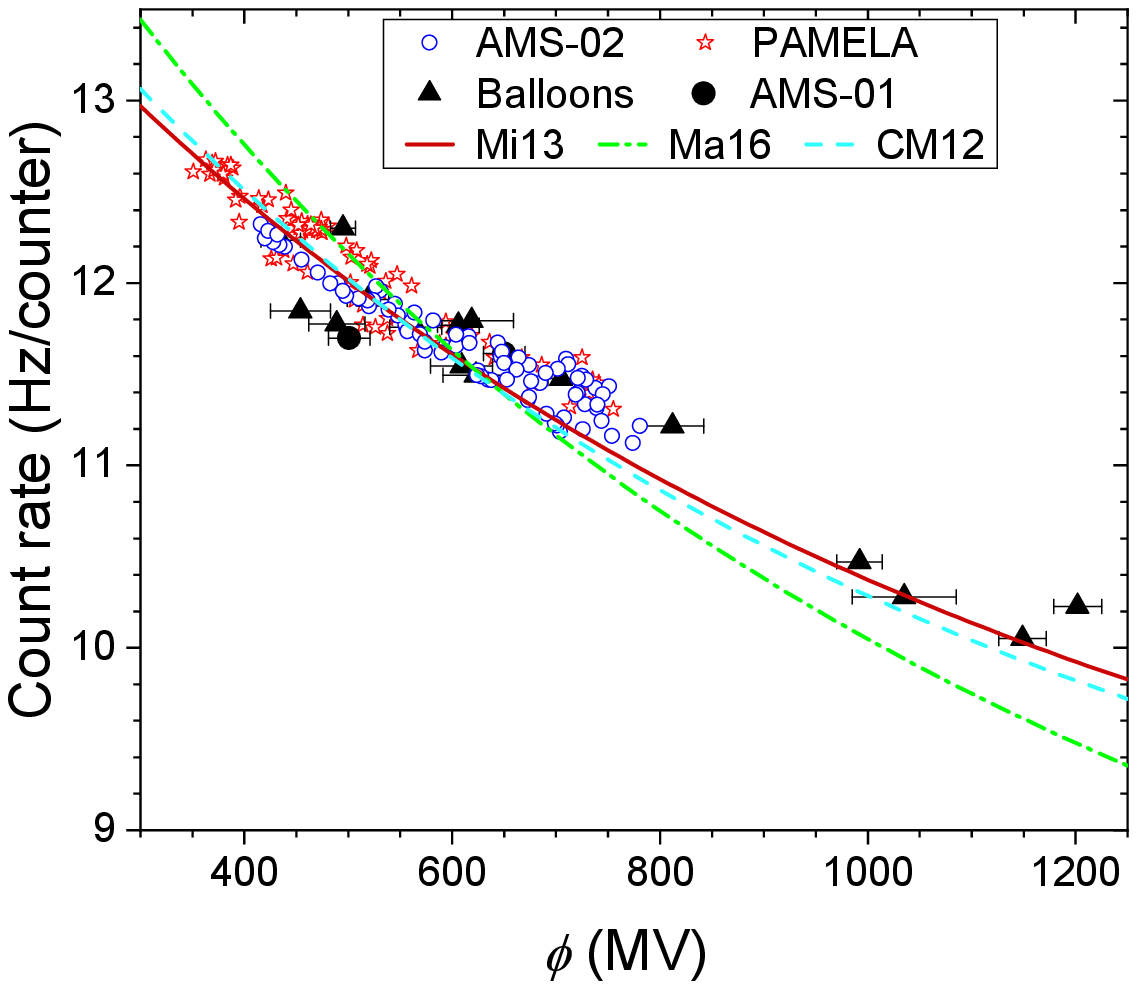}
	\caption{Dependence of the NM count rates of Oulu NM on the modulation potential, as
 obtained from directly measured spectra by space- and balloon-borne instruments.
 The points corresponding to the AMS-02 data were obtained here, while the others were adopted from elsewhere \citep{usoskin_gil_17,koldobsky18}.
 Theoretically expected dependencies (see text) for the Mi13, Ma16 and CM12 yield functions scaled with the $\kappa-$factors from Table~\ref{Tab:NM}
  are denoted by curves (the CD00-based curve is not shown as it is nearly identical to the Ma16 one).}
	\label{Fig:phi_n}
\end{figure}

We show in Figure~\ref{Fig:phi_n} the relation between the modulation potential $\phi$, obtained as the best-fit to the experimental data,
 and the NM count rate of Oulu NM for the same period.
The points for balloons, AMS-01 and PAMELA were adopted from elsewhere \citep{usoskin_gil_17,koldobsky18}, while
 the AMS-02 based points were obtained here.
On the same plot, we show the theoretically expected relations, calculated using the force-field formalism for
 the yield functions and reduced to the realistic count rate of Oulu NM by applying the scaling factor from Table~\ref{Tab:NM}.
Heavier GCR species were considered by scaling helium with the energy-dependent factor shown in Figure~\ref{Fig:RR}.

\begin{figure}
	\centering
	\includegraphics[width=\columnwidth]{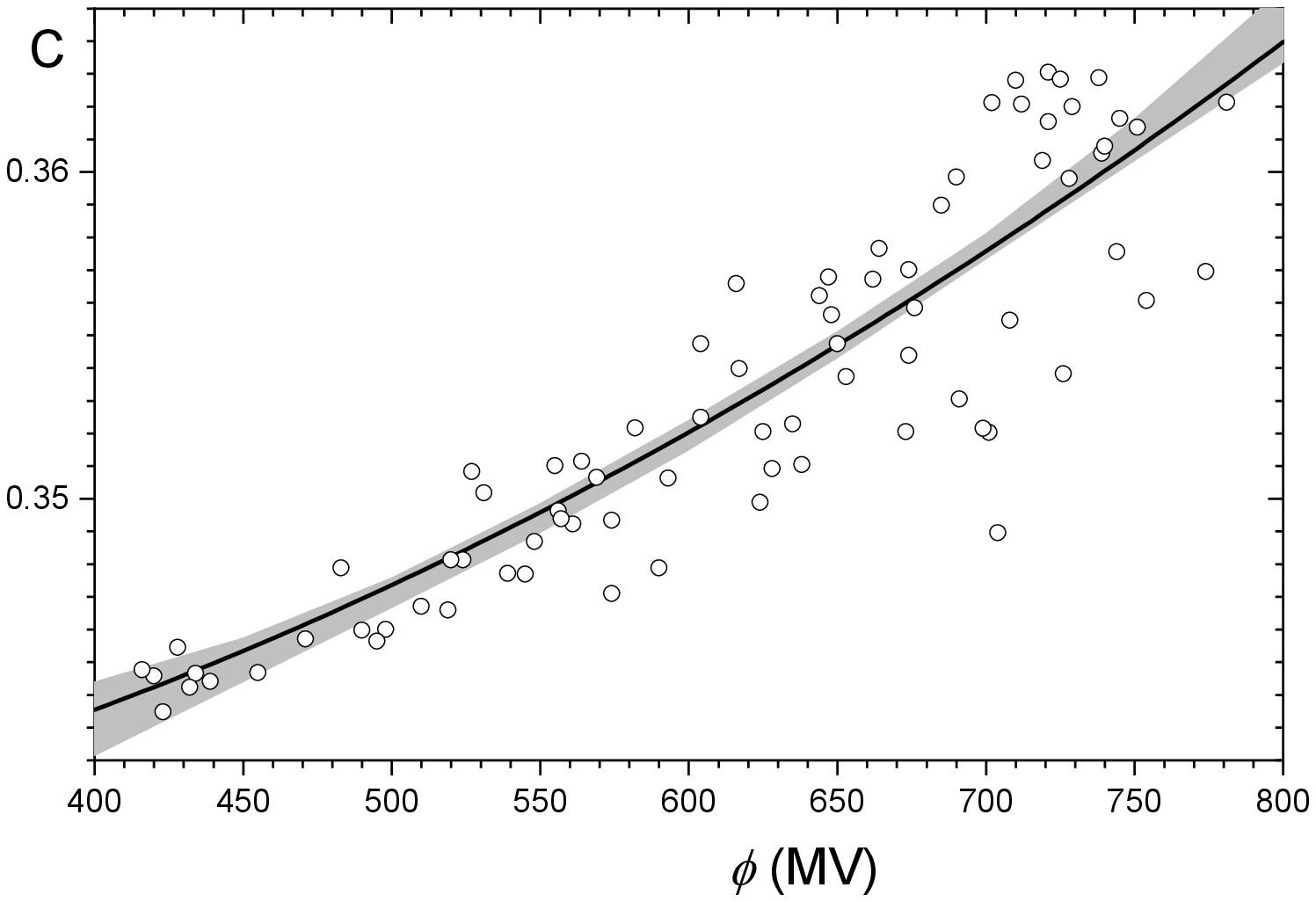}
	\caption{Dependence of the ratio $C=N_{\rm h, AMS}(\phi)/N_{\rm h, mod}$ on the modulation potential $\phi_{\rm p}$,
 where $N_{\rm h, AMS}$ is the contribution of heavier ($Z>1$) species to a polar sea-level NM directly computed from AMS-02 data
 (see Section~\ref{scale_factor}), while $N_{\rm h, mod}$ is that but calculated traditionally assuming that the LIS of heavier species is equal
 to that of protons (for the same energy per nucleon) and modulated using the force-field model with the modulation potential defined for
 protons $\phi_{\rm p}$.
 The dependence is shown for the Mi13 yield function.
 Points correspond to the individual BRs (see Table~\ref{Tab:phi}), the thick curve with grey shading represents
 the best-fit parabola (see text) with the 68\% confidence interval.
 If the standard assumption was correct, the dependence would have been a flat line at $C=0.3$.}
	\label{Fig:C}
\end{figure}

Here we also checked the standard assumption \citep[e.g.,][]{webber03,usoskin_Phi_05} that all heavier-than-proton
 species can be described by the same model as protons (local interstellar spectra are identical and only scaled, and the modulation
 is described by the force-field model with the same modulation potential).
The ratio of the directly computed (by applying the AMS-02 data as described above) response of a standard sea-level polar NM due to
 ($Z>1$) GCR species, $N_{\rm h, AMS}$, to the corresponding response, $N_{\rm h, mod}$, calculated by applying the standard force-field
 approach, as a function of the modulation potential $\phi$, viz.  $C(\phi)=N_{\rm h, AMS}(\phi)/N_{\rm h, mod}(\phi)$,
 is shown in Figure~\ref{Fig:C} for individual BRs.
It can be approximated by a parabolic dependence,
 $C=a\cdot\phi^2 + b\cdot\phi + c$, where $\phi$ is given in MV.
The best-fit parameters for the Mi13 YF are $a=(4.3\pm 0.9)\cdot 10^{-8}$;
 $b=(4.4\pm 1)\cdot 10^{-5}$ (all the parameters are very tightly connected to each other so that $b=-1134.2\cdot a + 4.8\cdot 10^{-5}$);
 $c=0.337\pm 0.003$ ($c=3.109\cdot 10^{-5}\cdot a + 0.323$).
One can see that the earlier used coefficient 0.3 for heavier species (see Section~\ref{Sec:h}) is not correct, since the mean
 $\langle C\rangle$ obtained here for the AMS-02 dataset is 0.353.
Importantly, there is also a significant solar cycle dependence, suggesting that the assumption of a constant ratio may lead
 to a systematic bias.
Accordingly, the contribution of heavier species to the NM count rate, computed in the `standard' way, should be corrected
 for this.
Modelled curves in Figure~\ref{Fig:phi_n} were calculated using this corrected dependence.

One can see in Figure~\ref{Fig:phi_n} that all the experimental data points are more or less consistent with each other and with the model curve.
However, theoretical curves, while matching the overall level due to the direct normalization to data, have systematic differences
 in the slope of the relation.
Mi13 yield function works fairly well in the entire studied modulation range, even exceeding the one covered by AMS-02 data.
The CM12 yield function lies close to the Mi13 one but leads to somewhat larger uncertainties.
The Ma16 yield function agrees with data only for the moderate activity  $\phi\approx$600 MV, where it has been normalized to,
 but over/under-estimates the count rates by about 5\,\% for low/high activity periods, respectively, giving the full range of $\approx$ 10\%
  or $\pm$120 MV in $\phi$ units.
This is a result of the effect, discussed above, that the Ma16 yield function likely overestimates the contribution of low-energy particles to the NM count rate.
The CD00 yield function is very close to the Ma16 one and also heavily overestimates the contribution of low-energy cosmic rays.


\section{Conclusions}
\label{conc}

The newly published spectra of protons and helium measured directly by AMS-02 for the period 2011\,--\,2017 for individual
 27-day BR periods allowed us to perform the full calibration and verification of the ground-based NM detectors.
Here we have presented the result of calibration of seven stable near-level NMs (Inuvik, Apatity, Oulu, Newark, Moscow, Hermanus, Athens)
 and calculated their scaling factors (Table~\ref{Tab:NM}) accounting for their `non-ideality'.

We have tested four modern NM yield functions (Mi13, Ma16, CM12 and CD00), by comparing the calculated NM count rates, on the
 basis of the cosmic-ray spectra measured by AMS-02 with those actually recorded for the corresponding time intervals.
The Mi13 yield function was found to realistically represent the NM response to GCRs.
In contrast, Ma16 and CD00 ones tend to overestimate the NM sensitivity to low-rigidity ($<10$ GV) cosmic rays, leading to a
 possible bias, of the order of 5\,--\,10\%, for the 11-year cycle in GCR.
This effect may be important if these yield functions are applied to an analysis of ground level enhancements, caused by solar energetic particles
 with much softer spectrum than GCR.
In particular it may lead to a significant underestimate of fluxes of solar energetic particles as based on NM data.
Accordingly, these yield functions need a revision in the lower-rigidity part.
CM12 yield function also depicts a possible small overestimate of the low-energy sensitivity of a NM, leading to a few percent bias.
We recommend to use Mi13 yield function for quantitative analyses of NM data.
This result does not necessarily imply that all previous studies of solar particle events were invalid,
 but this issue will be addressed in a forthcoming work.

We have also checked the validity of the force-field approximation, often applied to parameterize the GCR spectrum,
 and found that it indeed provides a good parametrization for the directly measured proton spectra, within a
 10\% uncertainty.
The accuracy of the approximation is very good (within a few \%) for periods of low to moderate activity and slowly degrades
 to $\approx 10$\,\% for active periods.

The results of this work strengthen and validate the method of cosmic-ray variability analysis based on the NM data and yield-function
 formalism, and improves its accuracy.

\acknowledgments
Data of AMS-02 were obtained from  NASA SPDF database
\url{http://www.spase-group.org/registry/render?f=yes&id=spase://VSPO/NumericalData/ISS/AMS-02/P27D}.
Data of NMs count rates were obtained from \url{http://cosmicrays.oulu.fi} (Oulu NM), \url{http://pgia.ru/CosmicRay/} (Apatity),
 Neutron Monitor Database (NMDB) and IZMIRAN Cosmic Ray data\-base (\url{http://cr0.izmiran.ru/common/links.htm}).
NMDB data\-base (\url{www.nmdb.eu)}, founded under the European Union's FP7 programme (contract no. 213007), is not responsible
 for the data quality.
Balloon and AMS-01 data were obtained from SSDC database (\url{https://tools.ssdc.asi.it/CosmicRays/}).
PIs and teams of all the balloon- and space-borne experiments as well as ground-based neutron monitors whose data
 were used here, are gratefully acknowledged.
This work was partially supported by the ReSoLVE Centre of Excellence (Academy of Finland, project no. 272157),
 by National Science Foundation Career Award under grant (NSF AGS-1455202); Wyle Laboratories, Inc. under grant (NAS 9-02078);
 NASA under grant (17-SDMSS17-0012), and by the Russian Foundation for Basic Research (grant 18-32-00062) and MEPhI Academic
 Excellence Project (contract 02.a03.21.0005)


%
%

\begin{thebibliography}{}

\bibitem [\protect \citeauthoryear {%
{Abunin}%
, {Pletnikov}%
, {Shchepetov}%
\BCBL {}\ \BBA {} {Yanke}%
}{%
{Abunin}%
\ \protect \BOthers {.}}{%
{\protect \APACyear {2011}}%
}]{%
abunin11}
\APACinsertmetastar {%
abunin11}%
\begin{APACrefauthors}%
{Abunin}, A\BPBI A.%
, {Pletnikov}, E\BPBI V.%
, {Shchepetov}, A\BPBI L.%
\BCBL {}\ \BBA {} {Yanke}, V\BPBI G.%
\end{APACrefauthors}%
\unskip\
\newblock
\APACrefYearMonthDay{2011}{}{}.
\newblock
{\BBOQ}\APACrefatitle {{Efficiency of detection for neutron detectors with
  different geometries}} {{Efficiency of detection for neutron detectors with
  different geometries}}.{\BBCQ}
\newblock
\APACjournalVolNumPages{Bull. Russian Acad. Science, Phys.}{75}{}{866-868}.
\newblock
\begin{APACrefDOI} \doi{10.3103/S1062873811060037} \end{APACrefDOI}
\PrintBackRefs{\CurrentBib}

\bibitem [\protect \citeauthoryear {%
Adriani%
\ \protect \BOthers {.}}{%
Adriani%
\ \protect \BOthers {.}}{%
{\protect \APACyear {2017}}%
}]{%
adriani17}
\APACinsertmetastar {%
adriani17}%
\begin{APACrefauthors}%
Adriani, O.%
, Barbarino, G.%
, Bazilevskaya, G.%
, Bellotti, R.%
, Boezio, M.%
, Bogomolov, E.%
\BDBL {}Zampa, N.%
\end{APACrefauthors}%
\unskip\
\newblock
\APACrefYearMonthDay{2017}{}{}.
\newblock
{\BBOQ}\APACrefatitle {{Ten years of PAMELA in space}} {{Ten years of PAMELA in
  space}}.{\BBCQ}
\newblock
\APACjournalVolNumPages{Rivista del Nuovo Cimento}{40}{10}{473-522}.
\newblock
\begin{APACrefDOI} \doi{10.1393/ncr/i2017-10140-x} \end{APACrefDOI}
\PrintBackRefs{\CurrentBib}

\bibitem [\protect \citeauthoryear {%
{Adriani}%
\ \protect \BOthers {.}}{%
{Adriani}%
\ \protect \BOthers {.}}{%
{\protect \APACyear {2014}}%
}]{%
adriani14}
\APACinsertmetastar {%
adriani14}%
\begin{APACrefauthors}%
{Adriani}, O.%
, {Barbarino}, G\BPBI C.%
, {Bazilevskaya}, G\BPBI A.%
, {Bellotti}, R.%
, {Boezio}, M.%
, {Bogomolov}, E\BPBI A.%
\BDBL {}{Zverev}, V\BPBI G.%
\end{APACrefauthors}%
\unskip\
\newblock
\APACrefYearMonthDay{2014}{}{}.
\newblock
{\BBOQ}\APACrefatitle {{The PAMELA Mission: Heralding a new era in precision
  cosmic ray physics}} {{The PAMELA Mission: Heralding a new era in precision
  cosmic ray physics}}.{\BBCQ}
\newblock
\APACjournalVolNumPages{Phys. Rep.}{544}{}{323-370}.
\newblock
\begin{APACrefDOI} \doi{10.1016/j.physrep.2014.06.003} \end{APACrefDOI}
\PrintBackRefs{\CurrentBib}

\bibitem [\protect \citeauthoryear {%
Aguilar%
\ \protect \BOthers {.}}{%
Aguilar%
\ \protect \BOthers {.}}{%
{\protect \APACyear {2015}}%
{\protect \APACexlab {{\protect \BCnt {1}}}}}]{%
aguilar_he_15}
\APACinsertmetastar {%
aguilar_he_15}%
\begin{APACrefauthors}%
Aguilar, M.%
, Aisa, D.%
, Alpat, B.%
, Alvino, A.%
, Ambrosi, G.%
, Andeen, K.%
\BDBL {}Zuccon, P.%
\end{APACrefauthors}%
\unskip\
\newblock
\APACrefYearMonthDay{2015{\protect \BCnt {1}}}{Nov}{}.
\newblock
{\BBOQ}\APACrefatitle {Precision Measurement of the Helium Flux in Primary
  Cosmic Rays of Rigidities 1.9 GV to 3 TV with the Alpha Magnetic Spectrometer
  on the International Space Station} {Precision measurement of the helium flux
  in primary cosmic rays of rigidities 1.9 gv to 3 tv with the alpha magnetic
  spectrometer on the international space station}.{\BBCQ}
\newblock
\APACjournalVolNumPages{Phys. Rev. Lett.}{115}{}{211101}.
\newblock
\begin{APACrefURL}
  \url{https://link.aps.org/doi/10.1103/PhysRevLett.115.211101}
  \end{APACrefURL}
\newblock
\begin{APACrefDOI} \doi{10.1103/PhysRevLett.115.211101} \end{APACrefDOI}
\PrintBackRefs{\CurrentBib}

\bibitem [\protect \citeauthoryear {%
Aguilar%
\ \protect \BOthers {.}}{%
Aguilar%
\ \protect \BOthers {.}}{%
{\protect \APACyear {2015}}%
{\protect \APACexlab {{\protect \BCnt {2}}}}}]{%
aguilar15}
\APACinsertmetastar {%
aguilar15}%
\begin{APACrefauthors}%
Aguilar, M.%
, Aisa, D.%
, Alpat, B.%
, Alvino, A.%
, Ambrosi, G.%
, Andeen, K.%
\BDBL {}Zurbach, C.%
\end{APACrefauthors}%
\unskip\
\newblock
\APACrefYearMonthDay{2015{\protect \BCnt {2}}}{Apr}{}.
\newblock
{\BBOQ}\APACrefatitle {Precision Measurement of the Proton Flux in Primary
  Cosmic Rays from Rigidity 1 GV to 1.8 TV with the Alpha Magnetic Spectrometer
  on the International Space Station} {Precision measurement of the proton flux
  in primary cosmic rays from rigidity 1 gv to 1.8 tv with the alpha magnetic
  spectrometer on the international space station}.{\BBCQ}
\newblock
\APACjournalVolNumPages{Phys. Rev. Lett.}{114}{}{171103}.
\newblock
\begin{APACrefURL}
  \url{https://link.aps.org/doi/10.1103/PhysRevLett.114.171103}
  \end{APACrefURL}
\newblock
\begin{APACrefDOI} \doi{10.1103/PhysRevLett.114.171103} \end{APACrefDOI}
\PrintBackRefs{\CurrentBib}

\bibitem [\protect \citeauthoryear {%
{Aguilar}%
\ \protect \BOthers {.}}{%
{Aguilar}%
\ \protect \BOthers {.}}{%
{\protect \APACyear {2013}}%
}]{%
aguilar_AMS_13}
\APACinsertmetastar {%
aguilar_AMS_13}%
\begin{APACrefauthors}%
{Aguilar}, M.%
, {Alberti}, G.%
, {Alpat}, B.%
, {Alvino}, A.%
, {Ambrosi}, G.%
, {Andeen}, K.%
\BDBL {}et al.%
\end{APACrefauthors}%
\unskip\
\newblock
\APACrefYearMonthDay{2013}{}{}.
\newblock
{\BBOQ}\APACrefatitle {{First result from the Alpha Magnetic Spectrometer on
  the International Space Station: precision measurement of the positron
  fraction in primary cosmic rays of 0.5--350 GeV}} {{First result from the
  Alpha Magnetic Spectrometer on the International Space Station: precision
  measurement of the positron fraction in primary cosmic rays of 0.5--350
  GeV}}.{\BBCQ}
\newblock
\APACjournalVolNumPages{Phys. Rev. Lett.}{110}{14}{141102}.
\newblock
\begin{APACrefDOI} \doi{10.1103/PhysRevLett.110.141102} \end{APACrefDOI}
\PrintBackRefs{\CurrentBib}

\bibitem [\protect \citeauthoryear {%
Aguilar%
\ \protect \BOthers {.}}{%
Aguilar%
\ \protect \BOthers {.}}{%
{\protect \APACyear {2017}}%
}]{%
aguilar_AMS_17}
\APACinsertmetastar {%
aguilar_AMS_17}%
\begin{APACrefauthors}%
Aguilar, M.%
, Ali~Cavasonza, L.%
, Alpat, B.%
, Ambrosi, G.%
, Arruda, L.%
, Attig, N.%
\BDBL {}Zuccon, P.%
\end{APACrefauthors}%
\unskip\
\newblock
\APACrefYearMonthDay{2017}{}{}.
\newblock
{\BBOQ}\APACrefatitle {Observation of the Identical Rigidity Dependence of He,
  C, and O Cosmic Rays at High Rigidities by the Alpha Magnetic Spectrometer on
  the International Space Station} {Observation of the identical rigidity
  dependence of he, c, and o cosmic rays at high rigidities by the alpha
  magnetic spectrometer on the international space station}.{\BBCQ}
\newblock
\APACjournalVolNumPages{Phys. Rev. Lett.}{119}{}{251101}.
\newblock
\begin{APACrefURL}
  \url{https://link.aps.org/doi/10.1103/PhysRevLett.119.251101}
  \end{APACrefURL}
\newblock
\begin{APACrefDOI} \doi{10.1103/PhysRevLett.119.251101} \end{APACrefDOI}
\PrintBackRefs{\CurrentBib}

\bibitem [\protect \citeauthoryear {%
Aguilar%
, Ali~Cavasonza%
, Alpat%
\BCBL {}\ \protect \BOthers {.}}{%
Aguilar%
, Ali~Cavasonza%
, Alpat%
\BCBL {}\ \protect \BOthers {.}}{%
{\protect \APACyear {2018}}%
{\protect \APACexlab {{\protect \BCnt {1}}}}}]{%
aguilar_AMS_18}
\APACinsertmetastar {%
aguilar_AMS_18}%
\begin{APACrefauthors}%
Aguilar, M.%
, Ali~Cavasonza, L.%
, Alpat, B.%
, Ambrosi, G.%
, Arruda, L.%
, Attig, N.%
\BDBL {}Zuccon, P.%
\end{APACrefauthors}%
\unskip\
\newblock
\APACrefYearMonthDay{2018{\protect \BCnt {1}}}{}{}.
\newblock
{\BBOQ}\APACrefatitle {Observation of Fine Time Structures in the Cosmic Proton
  and Helium Fluxes with the Alpha Magnetic Spectrometer on the International
  Space Station} {Observation of fine time structures in the cosmic proton and
  helium fluxes with the alpha magnetic spectrometer on the international space
  station}.{\BBCQ}
\newblock
\APACjournalVolNumPages{Phys. Rev. Lett.}{121}{}{051101}.
\newblock
\begin{APACrefURL}
  \url{https://link.aps.org/doi/10.1103/PhysRevLett.121.051101}
  \end{APACrefURL}
\newblock
\begin{APACrefDOI} \doi{10.1103/PhysRevLett.121.051101} \end{APACrefDOI}
\PrintBackRefs{\CurrentBib}

\bibitem [\protect \citeauthoryear {%
Aguilar%
, Ali~Cavasonza%
, Alpat%
\BCBL {}\ \protect \BOthers {.}}{%
Aguilar%
, Ali~Cavasonza%
, Alpat%
\BCBL {}\ \protect \BOthers {.}}{%
{\protect \APACyear {2018}}%
{\protect \APACexlab {{\protect \BCnt {2}}}}}]{%
aguilar_AMS_18_N}
\APACinsertmetastar {%
aguilar_AMS_18_N}%
\begin{APACrefauthors}%
Aguilar, M.%
, Ali~Cavasonza, L.%
, Alpat, B.%
, Ambrosi, G.%
, Arruda, L.%
, Attig, N.%
\BDBL {}Zuccon, P.%
\end{APACrefauthors}%
\unskip\
\newblock
\APACrefYearMonthDay{2018{\protect \BCnt {2}}}{}{}.
\newblock
{\BBOQ}\APACrefatitle {Precision Measurement of Cosmic-Ray Nitrogen and its
  Primary and Secondary Components with the Alpha Magnetic Spectrometer on the
  International Space Station} {Precision measurement of cosmic-ray nitrogen
  and its primary and secondary components with the alpha magnetic spectrometer
  on the international space station}.{\BBCQ}
\newblock
\APACjournalVolNumPages{Phys. Rev. Lett.}{121}{}{051103}.
\newblock
\begin{APACrefURL}
  \url{https://link.aps.org/doi/10.1103/PhysRevLett.121.051103}
  \end{APACrefURL}
\newblock
\begin{APACrefDOI} \doi{10.1103/PhysRevLett.121.051103} \end{APACrefDOI}
\PrintBackRefs{\CurrentBib}

\bibitem [\protect \citeauthoryear {%
Aguilar%
, Ali~Cavasonza%
, Ambrosi%
\BCBL {}\ \protect \BOthers {.}}{%
Aguilar%
, Ali~Cavasonza%
, Ambrosi%
\BCBL {}\ \protect \BOthers {.}}{%
{\protect \APACyear {2018}}%
}]{%
aguilar_AMS_18_sec}
\APACinsertmetastar {%
aguilar_AMS_18_sec}%
\begin{APACrefauthors}%
Aguilar, M.%
, Ali~Cavasonza, L.%
, Ambrosi, G.%
, Arruda, L.%
, Attig, N.%
, Aupetit, S.%
\BDBL {}Zuccon, P.%
\end{APACrefauthors}%
\unskip\
\newblock
\APACrefYearMonthDay{2018}{}{}.
\newblock
{\BBOQ}\APACrefatitle {Observation of New Properties of Secondary Cosmic Rays
  Lithium, Beryllium, and Boron by the Alpha Magnetic Spectrometer on the
  International Space Station} {Observation of new properties of secondary
  cosmic rays lithium, beryllium, and boron by the alpha magnetic spectrometer
  on the international space station}.{\BBCQ}
\newblock
\APACjournalVolNumPages{Phys. Rev. Lett.}{120}{}{021101}.
\newblock
\begin{APACrefURL}
  \url{https://link.aps.org/doi/10.1103/PhysRevLett.120.021101}
  \end{APACrefURL}
\newblock
\begin{APACrefDOI} \doi{10.1103/PhysRevLett.120.021101} \end{APACrefDOI}
\PrintBackRefs{\CurrentBib}

\bibitem [\protect \citeauthoryear {%
{Ahluwalia}%
\ \BBA {} {Ygbuhay}%
}{%
{Ahluwalia}%
\ \BBA {} {Ygbuhay}%
}{%
{\protect \APACyear {2013}}%
}]{%
ahluwalia13}
\APACinsertmetastar {%
ahluwalia13}%
\begin{APACrefauthors}%
{Ahluwalia}, H\BPBI S.%
\BCBT {}\ \BBA {} {Ygbuhay}, R\BPBI C.%
\end{APACrefauthors}%
\unskip\
\newblock
\APACrefYearMonthDay{2013}{}{}.
\newblock
{\BBOQ}\APACrefatitle {{Testing baseline stability of some neutron monitors in
  Europe, Africa, and Asia}} {{Testing baseline stability of some neutron
  monitors in Europe, Africa, and Asia}}.{\BBCQ}
\newblock
\APACjournalVolNumPages{Adv. Space Res.}{51}{}{1990-1995}.
\newblock
\begin{APACrefDOI} \doi{10.1016/j.asr.2013.01.014} \end{APACrefDOI}
\PrintBackRefs{\CurrentBib}

\bibitem [\protect \citeauthoryear {%
{Asvestari}%
, {Gil}%
, {Kovaltsov}%
\BCBL {}\ \BBA {} {Usoskin}%
}{%
{Asvestari}%
\ \protect \BOthers {.}}{%
{\protect \APACyear {2017}}%
}]{%
asvestari_JGR_17}
\APACinsertmetastar {%
asvestari_JGR_17}%
\begin{APACrefauthors}%
{Asvestari}, E.%
, {Gil}, A.%
, {Kovaltsov}, G\BPBI A.%
\BCBL {}\ \BBA {} {Usoskin}, I\BPBI G.%
\end{APACrefauthors}%
\unskip\
\newblock
\APACrefYearMonthDay{2017}{}{}.
\newblock
{\BBOQ}\APACrefatitle {{Neutron Monitors and Cosmogenic Isotopes as Cosmic Ray
  Energy-Integration Detectors: Effective Yield Functions, Effective Energy,
  and Its Dependence on the Local Interstellar Spectrum}} {{Neutron Monitors
  and Cosmogenic Isotopes as Cosmic Ray Energy-Integration Detectors: Effective
  Yield Functions, Effective Energy, and Its Dependence on the Local
  Interstellar Spectrum}}.{\BBCQ}
\newblock
\APACjournalVolNumPages{J. Geophys. Res. (Space Phys.)}{122}{}{9790-9802}.
\newblock
\begin{APACrefDOI} \doi{10.1002/2017JA024469} \end{APACrefDOI}
\PrintBackRefs{\CurrentBib}

\bibitem [\protect \citeauthoryear {%
{Belov}%
}{%
{Belov}%
}{%
{\protect \APACyear {2000}}%
}]{%
belov00}
\APACinsertmetastar {%
belov00}%
\begin{APACrefauthors}%
{Belov}, A.%
\end{APACrefauthors}%
\unskip\
\newblock
\APACrefYearMonthDay{2000}{}{}.
\newblock
{\BBOQ}\APACrefatitle {{Large Scale Modulation: View From the Earth}} {{Large
  Scale Modulation: View From the Earth}}.{\BBCQ}
\newblock
\APACjournalVolNumPages{Space Sci. Rev.}{93}{}{79-105}.
\newblock
\begin{APACrefDOI} \doi{10.1023/A:1026584109817} \end{APACrefDOI}
\PrintBackRefs{\CurrentBib}

\bibitem [\protect \citeauthoryear {%
Caballero-Lopez%
\ \BBA {} Moraal%
}{%
Caballero-Lopez%
\ \BBA {} Moraal%
}{%
{\protect \APACyear {2004}}%
}]{%
caballero04}
\APACinsertmetastar {%
caballero04}%
\begin{APACrefauthors}%
Caballero-Lopez, R.%
\BCBT {}\ \BBA {} Moraal, H.%
\end{APACrefauthors}%
\unskip\
\newblock
\APACrefYearMonthDay{2004}{}{}.
\newblock
{\BBOQ}\APACrefatitle {Limitations of the force field equation to describe
  cosmic ray modulation} {Limitations of the force field equation to describe
  cosmic ray modulation}.{\BBCQ}
\newblock
\APACjournalVolNumPages{J. Geophys. Res.}{109}{}{A01101}.
\newblock
\begin{APACrefDOI} \doi{10.1029/2003JA010098} \end{APACrefDOI}
\PrintBackRefs{\CurrentBib}

\bibitem [\protect \citeauthoryear {%
Caballero-Lopez%
\ \BBA {} Moraal%
}{%
Caballero-Lopez%
\ \BBA {} Moraal%
}{%
{\protect \APACyear {2012}}%
}]{%
caballero12}
\APACinsertmetastar {%
caballero12}%
\begin{APACrefauthors}%
Caballero-Lopez, R.%
\BCBT {}\ \BBA {} Moraal, H.%
\end{APACrefauthors}%
\unskip\
\newblock
\APACrefYearMonthDay{2012}{}{}.
\newblock
{\BBOQ}\APACrefatitle {Cosmic-Ray Yield and Response Functions in the
  Atmosphere} {Cosmic-ray yield and response functions in the
  atmosphere}.{\BBCQ}
\newblock
\APACjournalVolNumPages{J. Geophys. Res.}{117}{}{{A12103}}.
\newblock
\begin{APACrefDOI} \doi{10.1029/2012JA017794} \end{APACrefDOI}
\PrintBackRefs{\CurrentBib}

\bibitem [\protect \citeauthoryear {%
Clem%
\ \BBA {} Dorman%
}{%
Clem%
\ \BBA {} Dorman%
}{%
{\protect \APACyear {2000}}%
}]{%
clem00}
\APACinsertmetastar {%
clem00}%
\begin{APACrefauthors}%
Clem, J.%
\BCBT {}\ \BBA {} Dorman, L.%
\end{APACrefauthors}%
\unskip\
\newblock
\APACrefYearMonthDay{2000}{}{}.
\newblock
{\BBOQ}\APACrefatitle {Neutron Monitor Response Functions} {Neutron monitor
  response functions}.{\BBCQ}
\newblock
\APACjournalVolNumPages{Space Sci. Rev.}{93}{}{335--359}.
\newblock
\begin{APACrefDOI} \doi{10.1023/A:1026508915269} \end{APACrefDOI}
\PrintBackRefs{\CurrentBib}

\bibitem [\protect \citeauthoryear {%
Dorman%
}{%
Dorman%
}{%
{\protect \APACyear {2004}}%
}]{%
dorman04}
\APACinsertmetastar {%
dorman04}%
\begin{APACrefauthors}%
Dorman, L.%
\end{APACrefauthors}%
\unskip\
\newblock
\APACrefYear{2004}.
\newblock
\APACrefbtitle {Cosmic Rays in the Earth's Atmosphere and Underground} {Cosmic
  rays in the earth's atmosphere and underground}.
\newblock
\APACaddressPublisher{Dordrecht}{Kluwer Academic Publishers}.
\PrintBackRefs{\CurrentBib}

\bibitem [\protect \citeauthoryear {%
{Gieseler}%
, {Heber}%
\BCBL {}\ \BBA {} {Herbst}%
}{%
{Gieseler}%
\ \protect \BOthers {.}}{%
{\protect \APACyear {2017}}%
}]{%
geiseler18}
\APACinsertmetastar {%
geiseler18}%
\begin{APACrefauthors}%
{Gieseler}, J.%
, {Heber}, B.%
\BCBL {}\ \BBA {} {Herbst}, K.%
\end{APACrefauthors}%
\unskip\
\newblock
\APACrefYearMonthDay{2017}{}{}.
\newblock
{\BBOQ}\APACrefatitle {{An Empirical Modification of the Force Field Approach
  to Describe the Modulation of Galactic Cosmic Rays Close to Earth in a Broad
  Range of Rigidities}} {{An Empirical Modification of the Force Field Approach
  to Describe the Modulation of Galactic Cosmic Rays Close to Earth in a Broad
  Range of Rigidities}}.{\BBCQ}
\newblock
\APACjournalVolNumPages{J. Geophys. Res. (Space Phys.)}{122}{}{10}.
\newblock
\begin{APACrefDOI} \doi{10.1002/2017JA024763} \end{APACrefDOI}
\PrintBackRefs{\CurrentBib}

\bibitem [\protect \citeauthoryear {%
{Gil}%
\ \protect \BOthers {.}}{%
{Gil}%
\ \protect \BOthers {.}}{%
{\protect \APACyear {2015}}%
}]{%
gil15}
\APACinsertmetastar {%
gil15}%
\begin{APACrefauthors}%
{Gil}, A.%
, {Usoskin}, I\BPBI G.%
, {Kovaltsov}, G\BPBI A.%
, {Mishev}, A\BPBI L.%
, {Corti}, C.%
\BCBL {}\ \BBA {} {Bindi}, V.%
\end{APACrefauthors}%
\unskip\
\newblock
\APACrefYearMonthDay{2015}{}{}.
\newblock
{\BBOQ}\APACrefatitle {{Can we properly model the neutron monitor count rate?}}
  {{Can we properly model the neutron monitor count rate?}}{\BBCQ}
\newblock
\APACjournalVolNumPages{J. Geophys. Res.}{120}{}{7172-7178}.
\newblock
\begin{APACrefDOI} \doi{10.1002/2015JA021654} \end{APACrefDOI}
\PrintBackRefs{\CurrentBib}

\bibitem [\protect \citeauthoryear {%
Gleeson%
\ \BBA {} Axford%
}{%
Gleeson%
\ \BBA {} Axford%
}{%
{\protect \APACyear {1968}}%
}]{%
gleeson68}
\APACinsertmetastar {%
gleeson68}%
\begin{APACrefauthors}%
Gleeson, L.%
\BCBT {}\ \BBA {} Axford, W.%
\end{APACrefauthors}%
\unskip\
\newblock
\APACrefYearMonthDay{1968}{}{}.
\newblock
{\BBOQ}\APACrefatitle {Solar Modulation of Galactic Cosmic Rays} {Solar
  modulation of galactic cosmic rays}.{\BBCQ}
\newblock
\APACjournalVolNumPages{Astrophys. J.}{154}{}{1011--1026}.
\newblock
\begin{APACrefDOI} \doi{10.1086/149822} \end{APACrefDOI}
\PrintBackRefs{\CurrentBib}

\bibitem [\protect \citeauthoryear {%
{Herbst}%
\ \protect \BOthers {.}}{%
{Herbst}%
\ \protect \BOthers {.}}{%
{\protect \APACyear {2010}}%
}]{%
herbst10}
\APACinsertmetastar {%
herbst10}%
\begin{APACrefauthors}%
{Herbst}, K.%
, {Kopp}, A.%
, {Heber}, B.%
, {Steinhilber}, F.%
, {Fichtner}, H.%
, {Scherer}, K.%
\BCBL {}\ \BBA {} {Matthi{\"a}}, D.%
\end{APACrefauthors}%
\unskip\
\newblock
\APACrefYearMonthDay{2010}{}{}.
\newblock
{\BBOQ}\APACrefatitle {{On the importance of the local interstellar spectrum
  for the solar modulation parameter}} {{On the importance of the local
  interstellar spectrum for the solar modulation parameter}}.{\BBCQ}
\newblock
\APACjournalVolNumPages{J. Geophys. Res.}{115}{}{D00I20}.
\newblock
\begin{APACrefDOI} \doi{10.1029/2009JD012557} \end{APACrefDOI}
\PrintBackRefs{\CurrentBib}

\bibitem [\protect \citeauthoryear {%
{Koldobskiy}%
, {Kovaltsov}%
\BCBL {}\ \BBA {} {Usoskin}%
}{%
{Koldobskiy}%
\ \protect \BOthers {.}}{%
{\protect \APACyear {2018}}%
}]{%
koldobsky18}
\APACinsertmetastar {%
koldobsky18}%
\begin{APACrefauthors}%
{Koldobskiy}, S\BPBI A.%
, {Kovaltsov}, G\BPBI A.%
\BCBL {}\ \BBA {} {Usoskin}, I\BPBI G.%
\end{APACrefauthors}%
\unskip\
\newblock
\APACrefYearMonthDay{2018}{}{}.
\newblock
{\BBOQ}\APACrefatitle {{A Solar Cycle of Cosmic Ray Fluxes for 2006-2014:
  Comparison between PAMELA and Neutron Monitors}} {{A Solar Cycle of Cosmic
  Ray Fluxes for 2006-2014: Comparison between PAMELA and Neutron
  Monitors}}.{\BBCQ}
\newblock
\APACjournalVolNumPages{J. Geophys. Res. (Space Phys.)}{123}{}{4479-4487}.
\newblock
\begin{APACrefDOI} \doi{10.1029/2018JA025516} \end{APACrefDOI}
\PrintBackRefs{\CurrentBib}

\bibitem [\protect \citeauthoryear {%
{Mangeard}%
\ \protect \BOthers {.}}{%
{Mangeard}%
\ \protect \BOthers {.}}{%
{\protect \APACyear {2018}}%
}]{%
mangeard18}
\APACinsertmetastar {%
mangeard18}%
\begin{APACrefauthors}%
{Mangeard}, P\BHBI S.%
, {Clem}, J.%
, {Evenson}, P.%
, {Pyle}, R.%
, {Mitthumsiri}, W.%
, {Ruffolo}, D.%
\BDBL {}{Nutaro}, T.%
\end{APACrefauthors}%
\unskip\
\newblock
\APACrefYearMonthDay{2018}{}{}.
\newblock
{\BBOQ}\APACrefatitle {{Distinct Pattern of Solar Modulation of Galactic Cosmic
  Rays above a High Geomagnetic Cutoff Rigidity}} {{Distinct Pattern of Solar
  Modulation of Galactic Cosmic Rays above a High Geomagnetic Cutoff
  Rigidity}}.{\BBCQ}
\newblock
\APACjournalVolNumPages{Astrophys. J.}{858}{}{43}.
\newblock
\begin{APACrefDOI} \doi{10.3847/1538-4357/aabd3c} \end{APACrefDOI}
\PrintBackRefs{\CurrentBib}

\bibitem [\protect \citeauthoryear {%
{Mangeard}%
\ \protect \BOthers {.}}{%
{Mangeard}%
\ \protect \BOthers {.}}{%
{\protect \APACyear {2016}}%
}]{%
mangeard2}
\APACinsertmetastar {%
mangeard2}%
\begin{APACrefauthors}%
{Mangeard}, P\BHBI S.%
, {Ruffolo}, D.%
, {S{\'a}iz}, A.%
, {Nuntiyakul}, W.%
, {Bieber}, J\BPBI W.%
, {Clem}, J.%
\BDBL {}{Humble}, J\BPBI E.%
\end{APACrefauthors}%
\unskip\
\newblock
\APACrefYearMonthDay{2016}{}{}.
\newblock
{\BBOQ}\APACrefatitle {{Dependence of the neutron monitor count rate and time
  delay distribution on the rigidity spectrum of primary cosmic rays}}
  {{Dependence of the neutron monitor count rate and time delay distribution on
  the rigidity spectrum of primary cosmic rays}}.{\BBCQ}
\newblock
\APACjournalVolNumPages{J. Geophys. Res. (Space Phys.)}{121}{}{11620}.
\newblock
\begin{APACrefDOI} \doi{10.1002/2016JA023515} \end{APACrefDOI}
\PrintBackRefs{\CurrentBib}

\bibitem [\protect \citeauthoryear {%
{Maurin}%
, {Cheminet}%
, {Derome}%
, {Ghelfi}%
\BCBL {}\ \BBA {} {Hubert}%
}{%
{Maurin}%
\ \protect \BOthers {.}}{%
{\protect \APACyear {2015}}%
}]{%
maurin15}
\APACinsertmetastar {%
maurin15}%
\begin{APACrefauthors}%
{Maurin}, D.%
, {Cheminet}, A.%
, {Derome}, L.%
, {Ghelfi}, A.%
\BCBL {}\ \BBA {} {Hubert}, G.%
\end{APACrefauthors}%
\unskip\
\newblock
\APACrefYearMonthDay{2015}{}{}.
\newblock
{\BBOQ}\APACrefatitle {{Neutron monitors and muon detectors for solar
  modulation studies: Interstellar flux, yield function, and assessment of
  critical parameters in count rate calculations}} {{Neutron monitors and muon
  detectors for solar modulation studies: Interstellar flux, yield function,
  and assessment of critical parameters in count rate calculations}}.{\BBCQ}
\newblock
\APACjournalVolNumPages{Adv. Space Res.}{55}{}{363-389}.
\newblock
\begin{APACrefDOI} \doi{10.1016/j.asr.2014.06.021} \end{APACrefDOI}
\PrintBackRefs{\CurrentBib}

\bibitem [\protect \citeauthoryear {%
{Mishev}%
, {Usoskin}%
\BCBL {}\ \BBA {} {Kovaltsov}%
}{%
{Mishev}%
\ \protect \BOthers {.}}{%
{\protect \APACyear {2013}}%
}]{%
mishev13}
\APACinsertmetastar {%
mishev13}%
\begin{APACrefauthors}%
{Mishev}, A.%
, {Usoskin}, I.%
\BCBL {}\ \BBA {} {Kovaltsov}, G.%
\end{APACrefauthors}%
\unskip\
\newblock
\APACrefYearMonthDay{2013}{}{}.
\newblock
{\BBOQ}\APACrefatitle {{Neutron monitor yield function: New improved
  computations}} {{Neutron monitor yield function: New improved
  computations}}.{\BBCQ}
\newblock
\APACjournalVolNumPages{J. Geophys. Res. (Space Phys.)}{118}{}{2783-2788}.
\newblock
\begin{APACrefDOI} \doi{10.1002/jgra.50325} \end{APACrefDOI}
\PrintBackRefs{\CurrentBib}

\bibitem [\protect \citeauthoryear {%
{Potgieter}%
}{%
{Potgieter}%
}{%
{\protect \APACyear {2013}}%
}]{%
potgieterLR}
\APACinsertmetastar {%
potgieterLR}%
\begin{APACrefauthors}%
{Potgieter}, M.%
\end{APACrefauthors}%
\unskip\
\newblock
\APACrefYearMonthDay{2013}{}{}.
\newblock
{\BBOQ}\APACrefatitle {{Solar Modulation of Cosmic Rays}} {{Solar Modulation of
  Cosmic Rays}}.{\BBCQ}
\newblock
\APACjournalVolNumPages{Living Rev. Solar Phys.}{10}{}{3}.
\newblock
\begin{APACrefDOI} \doi{10.12942/lrsp-2013-3} \end{APACrefDOI}
\PrintBackRefs{\CurrentBib}

\bibitem [\protect \citeauthoryear {%
{Shea}%
\ \BBA {} {Smart}%
}{%
{Shea}%
\ \BBA {} {Smart}%
}{%
{\protect \APACyear {2000}}%
}]{%
shea_SSR_00}
\APACinsertmetastar {%
shea_SSR_00}%
\begin{APACrefauthors}%
{Shea}, M\BPBI A.%
\BCBT {}\ \BBA {} {Smart}, D\BPBI F.%
\end{APACrefauthors}%
\unskip\
\newblock
\APACrefYearMonthDay{2000}{}{}.
\newblock
{\BBOQ}\APACrefatitle {{Fifty Years of Cosmic Radiation Data}} {{Fifty Years of
  Cosmic Radiation Data}}.{\BBCQ}
\newblock
\APACjournalVolNumPages{Space Sci. Rev.}{93}{}{229-262}.
\newblock
\begin{APACrefDOI} \doi{10.1023/A:1026500713452} \end{APACrefDOI}
\PrintBackRefs{\CurrentBib}

\bibitem [\protect \citeauthoryear {%
{Simpson}%
}{%
{Simpson}%
}{%
{\protect \APACyear {2000}}%
}]{%
simpson00}
\APACinsertmetastar {%
simpson00}%
\begin{APACrefauthors}%
{Simpson}, J\BPBI A.%
\end{APACrefauthors}%
\unskip\
\newblock
\APACrefYearMonthDay{2000}{}{}.
\newblock
{\BBOQ}\APACrefatitle {{The Cosmic Ray Nucleonic Component: The Invention and
  Scientific Uses of the Neutron Monitor}} {{The Cosmic Ray Nucleonic
  Component: The Invention and Scientific Uses of the Neutron Monitor}}.{\BBCQ}
\newblock
\APACjournalVolNumPages{Space Sci. Rev.}{93}{}{11-32}.
\newblock
\begin{APACrefDOI} \doi{10.1023/A:1026567706183} \end{APACrefDOI}
\PrintBackRefs{\CurrentBib}

\bibitem [\protect \citeauthoryear {%
{Tanabashi}%
\ \protect \BOthers {.}}{%
{Tanabashi}%
\ \protect \BOthers {.}}{%
{\protect \APACyear {2018}}%
}]{%
tanabashi18}
\APACinsertmetastar {%
tanabashi18}%
\begin{APACrefauthors}%
{Tanabashi}, M.%
, {Hagiwara}, K.%
, {Hikasa}, K.%
, {Nakamura}, K.%
, {Sumino}, Y.%
, {Takahashi}, F.%
\BDBL {}et al.%
\end{APACrefauthors}%
\unskip\
\newblock
\APACrefYearMonthDay{2018}{{\APACmonth{08}}}{}.
\newblock
{\BBOQ}\APACrefatitle {{Review of Particle Physics}} {{Review of Particle
  Physics}}.{\BBCQ}
\newblock
\APACjournalVolNumPages{Phys. Rev. D}{}{3}{030001}.
\newblock
\begin{APACrefDOI} \doi{10.1103/PhysRevD.98.030001} \end{APACrefDOI}
\PrintBackRefs{\CurrentBib}

\bibitem [\protect \citeauthoryear {%
{Tomassetti}%
\ \protect \BOthers {.}}{%
{Tomassetti}%
\ \protect \BOthers {.}}{%
{\protect \APACyear {2018}}%
}]{%
tomassetti18}
\APACinsertmetastar {%
tomassetti18}%
\begin{APACrefauthors}%
{Tomassetti}, N.%
, {Bar{\~a}o}, F.%
, {Bertucci}, B.%
, {Fiandrini}, E.%
, {Figueiredo}, J\BPBI L.%
, {Lousada}, J\BPBI B.%
\BCBL {}\ \BBA {} {Orcinha}, M.%
\end{APACrefauthors}%
\unskip\
\newblock
\APACrefYearMonthDay{2018}{}{}.
\newblock
{\BBOQ}\APACrefatitle {{Testing Diffusion of Cosmic Rays in the Heliosphere
  with Proton and Helium Data from AMS}} {{Testing Diffusion of Cosmic Rays in
  the Heliosphere with Proton and Helium Data from AMS}}.{\BBCQ}
\newblock
\APACjournalVolNumPages{Phys. Rev. Lett.}{121}{25}{251104}.
\newblock
\begin{APACrefDOI} \doi{10.1103/PhysRevLett.121.251104} \end{APACrefDOI}
\PrintBackRefs{\CurrentBib}

\bibitem [\protect \citeauthoryear {%
Usoskin%
, Alanko-Huotari%
, Kovaltsov%
\BCBL {}\ \BBA {} Mursula%
}{%
Usoskin%
\ \protect \BOthers {.}}{%
{\protect \APACyear {2005}}%
}]{%
usoskin_Phi_05}
\APACinsertmetastar {%
usoskin_Phi_05}%
\begin{APACrefauthors}%
Usoskin, I\BPBI G.%
, Alanko-Huotari, K.%
, Kovaltsov, G\BPBI A.%
\BCBL {}\ \BBA {} Mursula, K.%
\end{APACrefauthors}%
\unskip\
\newblock
\APACrefYearMonthDay{2005}{}{}.
\newblock
{\BBOQ}\APACrefatitle {{Heliospheric modulation of cosmic rays: Monthly
  reconstruction for 1951--2004}} {{Heliospheric modulation of cosmic rays:
  Monthly reconstruction for 1951--2004}}.{\BBCQ}
\newblock
\APACjournalVolNumPages{J. Geophys. Res.}{110}{}{{A12108}}.
\newblock
\begin{APACrefDOI} \doi{10.1029/2005JA011250} \end{APACrefDOI}
\PrintBackRefs{\CurrentBib}

\bibitem [\protect \citeauthoryear {%
{Usoskin}%
, {Gil}%
, {Kovaltsov}%
, {Mishev}%
\BCBL {}\ \BBA {} {Mikhailov}%
}{%
{Usoskin}%
\ \protect \BOthers {.}}{%
{\protect \APACyear {2017}}%
}]{%
usoskin_gil_17}
\APACinsertmetastar {%
usoskin_gil_17}%
\begin{APACrefauthors}%
{Usoskin}, I\BPBI G.%
, {Gil}, A.%
, {Kovaltsov}, G\BPBI A.%
, {Mishev}, A\BPBI L.%
\BCBL {}\ \BBA {} {Mikhailov}, V\BPBI V.%
\end{APACrefauthors}%
\unskip\
\newblock
\APACrefYearMonthDay{2017}{}{}.
\newblock
{\BBOQ}\APACrefatitle {{Heliospheric modulation of cosmic rays during the
  neutron monitor era: Calibration using PAMELA data for 2006-2010}}
  {{Heliospheric modulation of cosmic rays during the neutron monitor era:
  Calibration using PAMELA data for 2006-2010}}.{\BBCQ}
\newblock
\APACjournalVolNumPages{J. Geophys. Res. (Space Phys.)}{122}{}{3875-3887}.
\newblock
\begin{APACrefDOI} \doi{10.1002/2016JA023819} \end{APACrefDOI}
\PrintBackRefs{\CurrentBib}

\bibitem [\protect \citeauthoryear {%
{Vainio}%
\ \protect \BOthers {.}}{%
{Vainio}%
\ \protect \BOthers {.}}{%
{\protect \APACyear {2009}}%
}]{%
vainio09}
\APACinsertmetastar {%
vainio09}%
\begin{APACrefauthors}%
{Vainio}, R.%
, {Desorgher}, L.%
, {Heynderickx}, D.%
, {Storini}, M.%
, {Fl{\"u}ckiger}, E.%
, {Horne}, R\BPBI B.%
\BDBL {}{Usoskin}, I\BPBI G.%
\end{APACrefauthors}%
\unskip\
\newblock
\APACrefYearMonthDay{2009}{}{}.
\newblock
{\BBOQ}\APACrefatitle {{Dynamics of the Earth's particle radiation
  environment}} {{Dynamics of the Earth's particle radiation
  environment}}.{\BBCQ}
\newblock
\APACjournalVolNumPages{Space Sci. Rev.}{147}{}{187--231}.
\newblock
\begin{APACrefDOI} \doi{10.1007/s11214-009-9496-7} \end{APACrefDOI}
\PrintBackRefs{\CurrentBib}

\bibitem [\protect \citeauthoryear {%
{Vos}%
\ \BBA {} {Potgieter}%
}{%
{Vos}%
\ \BBA {} {Potgieter}%
}{%
{\protect \APACyear {2015}}%
}]{%
vos15}
\APACinsertmetastar {%
vos15}%
\begin{APACrefauthors}%
{Vos}, E\BPBI E.%
\BCBT {}\ \BBA {} {Potgieter}, M\BPBI S.%
\end{APACrefauthors}%
\unskip\
\newblock
\APACrefYearMonthDay{2015}{}{}.
\newblock
{\BBOQ}\APACrefatitle {{New Modeling of Galactic Proton Modulation during the
  Minimum of Solar Cycle 23/24}} {{New Modeling of Galactic Proton Modulation
  during the Minimum of Solar Cycle 23/24}}.{\BBCQ}
\newblock
\APACjournalVolNumPages{Astrophys. J.}{815}{}{119}.
\newblock
\begin{APACrefDOI} \doi{10.1088/0004-637X/815/2/119} \end{APACrefDOI}
\PrintBackRefs{\CurrentBib}

\bibitem [\protect \citeauthoryear {%
Webber%
\ \BBA {} Higbie%
}{%
Webber%
\ \BBA {} Higbie%
}{%
{\protect \APACyear {2003}}%
}]{%
webber03}
\APACinsertmetastar {%
webber03}%
\begin{APACrefauthors}%
Webber, W.%
\BCBT {}\ \BBA {} Higbie, P.%
\end{APACrefauthors}%
\unskip\
\newblock
\APACrefYearMonthDay{2003}{}{}.
\newblock
{\BBOQ}\APACrefatitle {Production of cosmogenic Be nuclei in the Earth's
  atmosphere by cosmic rays: Its dependence on solar modulation and the
  interstellar cosmic ray spectrum} {Production of cosmogenic be nuclei in the
  earth's atmosphere by cosmic rays: Its dependence on solar modulation and the
  interstellar cosmic ray spectrum}.{\BBCQ}
\newblock
\APACjournalVolNumPages{J. Geophys. Res.}{108}{}{1355}.
\newblock
\begin{APACrefDOI} \doi{10.1029/2003JA009863} \end{APACrefDOI}
\PrintBackRefs{\CurrentBib}

\end{thebibliography}

\end{document}